\documentclass[english]{aghengthesis}
\usepackage[utf8]{inputenc}
\usepackage{url}
\usepackage{placeins}
\usepackage{float}
\usepackage{subfigure}
\usepackage{tabularx}
\usepackage{ragged2e}
\usepackage{booktabs}
\usepackage{multirow}
\usepackage{grffile}
\usepackage{indentfirst}
\usepackage{caption}
\usepackage{listings}
\usepackage[ruled,linesnumbered,lined]{algorithm2e}
\usepackage[bookmarks=false]{hyperref}
\usepackage[nameinlink]{cleveref}
\usepackage[inkscapeformat=png]{svg}
\usepackage{graphicx}

\hypersetup{colorlinks,
  linkcolor=blue,
  citecolor=blue,
  urlcolor=blue}

\usepackage[
style=nature,
sorting=none,
isbn=false,
doi=true,
url=true,
backref=false,
backrefstyle=none,
maxnames=10,
giveninits=true,
abbreviate=true,
defernumbers=false,
backend=bibtex]{biblatex}
\usepackage{csquotes}
\DeclareQuoteStyle[quotes]{polish}
  {\quotedblbase}
  {\textquotedblright}
  [0.05em]
  {\quotesinglbase}
  {\fixligatures\textquoteright}
\DeclareQuoteAlias[quotes]{polish}{polish}

\usepackage[nottoc]{tocbibind}

\SetAlgorithmName{\LangAlgorithm}{\LangAlgorithmRef}{\LangListOfAlgorithms}

\newcolumntype{Y}{>{\small\centering\arraybackslash}X}
\author{Michał Szafarczyk, Piotr Ludynia, Przemysław Kukla}

\titlePL{Biblioteka do efektywnego obliczania fingerprintów molekularnych w języku Python}
\titleEN{A Python library for efficient computation of molecular fingerprints}

\fieldofstudy{Informatyka}

\supervisor{dr inż.\ Wojciech Czech}

\date{\the\year}

\begin{document}

\maketitle

\section*{Acknowledgments}
We express our heartfelt gratitude to Dr. Wojciech Czech for his dedicated supervision of our project. We appreciate the insightful subject he proposed for this thesis and the valuable feedback he provided. Special thanks go to Jakub Adamczyk for his unwavering support and expertise in machine learning and chemoinformatics. His consistent guidance and feedback played a crucial role in helping us navigate challenges and stay motivated throughout the project. Additionally, we extend our thanks to AGH University of Science and Technology, especially the Faculty of Computer Science, for their assistance with organizational matters.

\clearpage

\section*{Abstract}
Machine learning solutions are very popular in the field of chemoinformatics,  where they have numerous applications, such as novel drug discovery or molecular property prediction. Molecular fingerprints are algorithms commonly used for vectorizing chemical molecules as a part of preprocessing in this kind of solution. However, despite their popularity, there are no libraries that implement them efficiently for large datasets, utilizing modern, multicore architectures. On top of that, most of them do not provide the user with an intuitive interface, or one that would be compatible with other machine learning tools.

In this project, we created a Python library that computes molecular fingerprints efficiently and delivers an interface that is comprehensive and enables the user to easily incorporate the library into their existing machine learning workflow. The library enables the user to perform computation on large datasets using parallelism. Because of that, it is possible to perform such tasks as hyperparameter tuning in a reasonable time. We describe tools used in implementation of the library and asses its time performance on example benchmark datasets. Additionally, we show that using molecular fingerprints we can achieve results comparable to state-of-the-art ML solutions even with very simple models.

\tableofcontents

%%%%%%%%%%%%%%%%%%%%%%%%%%%%%%%%%%%%%%%%%%%%%%%%%%%%%%%%%%%%%%%%%%%%%%%%%%%%%%%
\chapter{\ChapterTitleProjectVision}
\label{sec:cel-wizja}

\noindent In this chapter, we describe the outline of our project and provide a few general remarks from the technical perspective.

%%%%%%%%%%%%%%%%%%%%%%%%%%%%%%%%%%%%%%%%%%%%%%%%%%%%%%%%%%%%%%%%%%%%%%%%%%%%%%%
\section{Description of the problem domain}
\label{sec:problem-description}

Molecular fingerprints are the class of feature extraction algorithms that can be used to embed a molecule as a vector in a metric space \cite{nc-mfp}. This allows to compare molecules based on their semantic similarity,  e.g. their chemical function or physicochemical properties. Efficient computing of molecular fingerprints is crucial in many fields, including drug discovery \cite{mdl-keys}, materials science \cite{Parsaeifard_2021}, and chemical analysis \cite{YANG2022103356}. Nevertheless, the currently available Python solutions do not meet the performance needs of machine learning scientists and are not user-friendly, making them cumbersome to use or even inaccessible to those, who lack advanced programming or data analysis skills. Additionally, only a few libraries for computing molecular fingerprints exist for Python programming language, with the majority being short-lived projects that promptly ceased development or failed to keep up with current standards. For example, most libraries lack support for basic programming techniques for efficient computations, like parallel programming. Also, they lack an interface compatible with the \texttt{scikit-learn} library \cite{scikit-learn}, which is currently the industry standard. As a result, there is a need for a more efficient and accessible tool that can provide scientists in diverse fields with a robust tool for analyzing molecular structures.

%%%%%%%%%%%%%%%%%%%%%%%%%%%%%%%%%%%%%%%%%%%%%%%%%%%%%%%%%%%%%%%%%%%%%%%%%%%%%%%
\section{General project vision}
\label{sec:project-vision}

The main goal of this thesis project is to create a Python library which allows for easy computation of several known and widely used molecular fingerprints. The library should be intuitive and easy to use. It should also be efficient on all layers, and use parallelism.  Additionally, its interface will be based on that of scikit-learn and other data science libraries to ensure compatibility and convenience for users already familiar with similar software. The final product will allow scientists to fully utilize the potential of molecular fingerprints in machine learning and allow for easy and efficient development of machine learning methods in the fields of chemistry and chemoinformatics.

In a molecular dataset, consisting of many chemical compounds, molecules can be processed into fingerprints independently, so the task is naturally parallel \cite{embarrassingly-parallel}, and we will be able to fully use modern multicore CPU architectures by utilizing the parallel computation. The optimization process will also take into account the algorithms themselves, e.g. by using the best Python code practices and source code optimizations. At first, Python language might not strike as the most effective language for fast computation. However, the heaviest part of the algorithm — the fingerprint computation itself — is already calculated with help of more effective languages such as C++. Python is our tool to implement the parallelism and user-friendly interfaces because of its wide use and popularity in modern machine learning (ML).

%%%%%%%%%%%%%%%%%%%%%%%%%%%%%%%%%%%%%%%%%%%%%%%%%%%%%%%%%%%%%%%%%%%%%%%%%%%%%%%

\section{The Scope of work}
\label{sec:scope-of-the-work}
This library will include multiple well-known algorithms for molecular fingerprints, such as ECFP \cite{ecfp}, MACCS Keys \cite{mdl-keys} or Atom Pair \cite{atom-pair}. However, we do not include pretrained deep learning models in this group, like for example Graphormer \cite{graphformer}, 1D CNNs \cite{baptista-eval-mol-rep} or language models like FP-BERT \cite{fp-bert}. The reason is that pretraining such models for out-of-the-box featurization of molecules is a very recent development, not yet well understood and not widely used by computational chemists. Additionally, it has been shown \cite{moleculenet, zagidullin-analysis, stepisnik-comparison, TDC} that traditional molecular fingerprints are fast to compute and give excellent results, often outperforming neural networks on various tasks.

%%%%%%%%%%%%%%%%%%%%%%%%%%%%%%%%%%%%%%%%%%%%%%%%%%%%%%%%%%%%%%%%%%%%%%%%%%%%%%%
\section{Other competitive solutions}
\label{sec:competition}

\begin{itemize}
\item RDKit \cite{rdkit} — arguably the most popular open-source chemoinformatics library. It provides a variety of tools for handling molecular data. It was originally developed by the pharmaceutical company Rational Discovery and is now maintained by a community of contributors. While RDKit provides a wide range of functionalities, the user interface can be complex and difficult for new users to navigate. It uses auto-generated documentation, which, combined with a mix of C++ and Python elements, makes it next to impossible to find anything. Moreover, its compatibility with other Python tools is limited by the poorly designed interface between Python and C++. Moreover, while it implements optimized calculation of fingerprints for singular molecules, it does not offer any parallelism for efficient processing of large datasets.

\item Scikit-chem \cite{scikit-chem} — a Python library that provides tools and algorithms for chemoinformatics and computational chemistry. It is built on top of popular data science libraries like NumPy \cite{harris2020array}, Pandas \cite{reback2020pandas}, and scikit-learn \cite{scikit-learn}, and offers a wide range of functionalities for working with molecular data. However, it is a general-purpose chemoinformatics library, and therefore it is not optimized specifically for molecular fingerprint computation. As its development was discontinued, it cannot be regarded as a working competitor.

\item Scikit-mol \cite{scikit-mol} — the intended usage of this library is to enable molecular vectorization directly in scikit-learn pipelines, so that the final model works directly on RDKit molecules or SMILES strings. It was initially created as a part of RDKit UGM 2022 hackathon. Realistically speaking, it is only a minimal wrapper for RDKit and has poor parallel computing — implemented only Python's multiprocessing without any optimization, e.g. for passing data between processes. The code quality is also not great. On the other hand, numerous notebooks with example usage is a considerable advantage. This solution can not be fully considered as a competitor, as it is no longer developed or updated.
\end{itemize}

%%%%%%%%%%%%%%%%%%%%%%%%%%%%%%%%%%%%%%%%%%%%%%%%%%%%%%%%%%%%%%%%%%%%%%%%%%%%%%%
\section{Useful technologies}
\label{sec:technologies}

\begin{itemize}
\item Joblib \cite{joblib} — the library for lightweight and efficient parallelism in Python. It is very effective at starting processes and passing objects between them. In particular, it works very well with NumPy matrices. In our project, it is used for easy and simple parallel computing of batched datasets of molecules.

\item NumPy \cite{harris2020array} — library for numerical and scientific computing. It provides an efficient implementation of multidimensional arrays based on C language with easy-to-use Python interface, along with mathematical functions for complex calculations. It enables vectorized operations, which significantly speeds up many tasks, and allows better memory management. We will use NumPy arrays as a default unit for storage and operation on molecules in our project.

\item SciPy \cite{2020SciPy-NMeth} — library providing tools for scientific computations, including differential equations, interpolation, and statistics. In particular, it offers an efficient implementation of sparse arrays and operations on them, which we use as one of possible output data formats along with NumPy arrays.

\item Scikit-learn \cite{scikit-learn} — the most popular Python library for machine learning that provides efficient tools for data mining and analysis. It is built on top of NumPy and SciPy, and provides a wide range of algorithms for classification, regression, clustering, and dimensionality reduction. It is widely regarded as a gold standard in terms of intuitive and easy-to-use programming interfaces for machine learning. For this reason, the whole ecosystem of supporting libraries keeps compatibility with scikit-learn interfaces. Such an integration is critically important for this project.

\item OGB — Open Graph Benchmark \cite{ogb} — a collection of benchmark datasets for molecular property prediction. OGB implements an interface for MoleculeNet \cite{moleculenet} datasets, which are the most commonly used datasets for benchmarking molecular chemistry algorithms.

\end{itemize}

%%%%%%%%%%%%%%%%%%%%%%%%%%%%%%%%%%%%%%%%%%%%%%%%%%%%%%%%%%%%%%%%%%%%%%%%%%%%%%%
\section{Risk assessment}
\label{sec:risks}

\begin{enumerate}
\item \textbf{Compatibility issues with existing libraries and software tools:} commonly used technologies are in constant development, and their APIs and functionalities might change during the time of creating the library. We need to consider, which versions of the most popular libraries should be initially supported. The team working on the project should be familiar with the technologies and how they change in time.
\item \textbf{Implementation mistakes:} this library needs to integrate various other projects and implement multiple algorithms. This includes parallel programming, which typically results in high complexity and more error-prone code. This can be avoided by designing good quality unit tests, with high test coverage.
\item \textbf{Low performance:} there is a chance that the use of Python programming language will not yield required speed. This can be mitigated with careful optimization and inclusion of language-specific constructs, as well as good choice of supporting computational libraries.
\item \textbf{Difficulty in readability:} if the source code is not readable, it might be hard for users to navigate around the library and therefore, not be able to use the library to its fullest capabilities. It is crucial that the code structure is designed with the proper care and thought of future users.
\end{enumerate}

%%%%%%%%%%%%%%%%%%%%%%%%%%%%%%%%%%%%%%%%%%%%%%%%%%%%%%%%%%%%%%%%%%%%%%%%%%%%%%%
\chapter{\ChapterTitleScope}
\label{sec:functional-scope}

\noindent In this chapter, we describe fingerprint calculation algorithms, which will be implemented in the project. We also list planned functionalities with their priority estimates, using MoSCoW method.

%%%%%%%%%%%%%%%%%%%%%%%%%%%%%%%%%%%%%%%%%%%%%%%%%%%%%%%%%%%%%%%%%%%%%%%%%%%%%%%
\section{Introduction to molecular fingerprints}
\label{sec:introduction-fp}

\textbf{Molecular fingerprints} are numerical representations of chemical structures as vectors in a metric space, originally designed to assist in chemical database substructure searching \cite{daylight}, but later used for various computational chemistry tasks, such as similarity searching \cite{daylight}, clustering \cite{clustering-fp}, and molecular property prediction \cite{fp-bert}.

\begin{figure}[h]\includegraphics[width=0.6\textwidth]{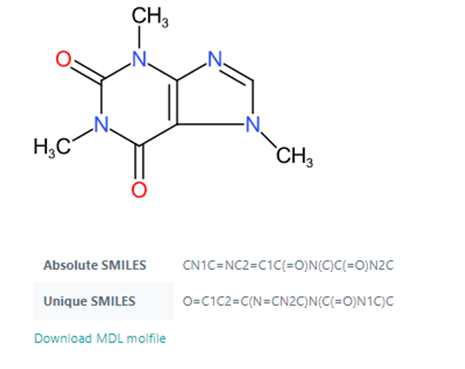}
    \centering
    \caption{Representation of a molecule in SMILES format \cite{smiles-language}.}
    \label{fig:smiles}
\end{figure}

Molecules are typically stored in string formats, like SMILES (shown in \cref{fig:smiles}) or FASTA (for proteins). Most ML methods are not able to use them in that raw form, so they need to be converted into graph representation. There are many ML methods that can use such graph representation to perform feature extraction, most prominently molecular fingerprints but also, e.g., Graph Neural Networks (GNNs) \cite{moleculenet}. 

Fingerprints allow us to work numerically with molecules. To calculate a fingerprint, we take a raw SMILES input, then work on the molecule's graph representation to utilize its topology, and return a multidimensional vector. In that form, we can use those feature vectors as an input, e.g. for machine learning (ML) algorithms. For most fingerprint algorithms, only the 2D structure is taken into account as it has reasonable computational complexity cost and for most purposes it is not necessary to include 3D (stereochemical) structure. This is especially true for small molecules, which are common, e.g., in drug discovery.

Generally speaking, we may divide molecular fingerprints into two groups or approaches \cite{mol-representations}. The first one acquires molecular features, often called \textbf{molecular descriptors}, by feature extraction. It uses predefined molecular substructures, which are designed by the experts in the field. It checks if those substructures exist in a given molecule, and this search forms a binary vector. The second one, often called \textbf{hashed fingerprints}, firstly extracts some structural or chemical properties of a molecule, which are later hashed to create a dictionary (hashmap) of molecular substructures. It is then converted into the final numerical vector of desired dimensionality, typically by a simple modulo operation. When it comes to the output format of fingerprint algorithms, there are also 2 options: dense or sparse matrices.

\begin{figure}[h]
    \includegraphics[width=0.8\textwidth]{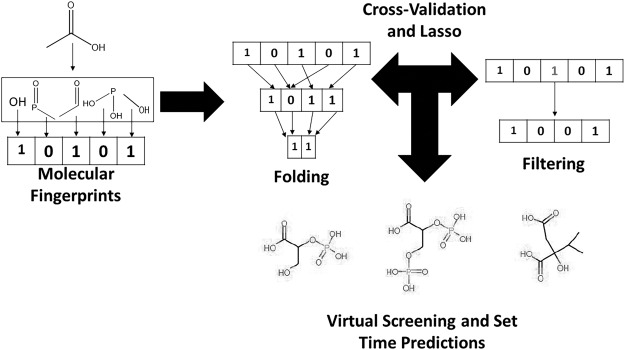}
    \centering
    \caption{The workflow for virtual screening \cite{accelerated-design-admixtures}.}
    \label{fig:folding}
\end{figure}

Additionally, the output vector from a hash fingerprint can be compressed further by a technique called \textbf{folding} \cite{daylight} (shown in \cref{fig:folding}). The folding process begins with a very high-dimensional fingerprint, with a size large enough to accurately represent any molecule we expect to encounter. The fingerprint is then folded: we divide it into two equal halves, then combine the two halves using a logical OR. Such compression results in some loss of information, but allows us to get a dense result and reduce the dimensionality.

For many fingerprinting algorithms, it is also possible to use a counted version of a fingerprint. It means, that when the same substructures occur multiple times, the algorithm counts each occurrence and includes that in the resulting vector, before hashing. However, this additional information is not always useful or beneficial, and therefore whether to use a binary or count version of a fingerprint is an important hyperparameter.

\begin{figure}[h]
    \includegraphics[width=0.7\textwidth]{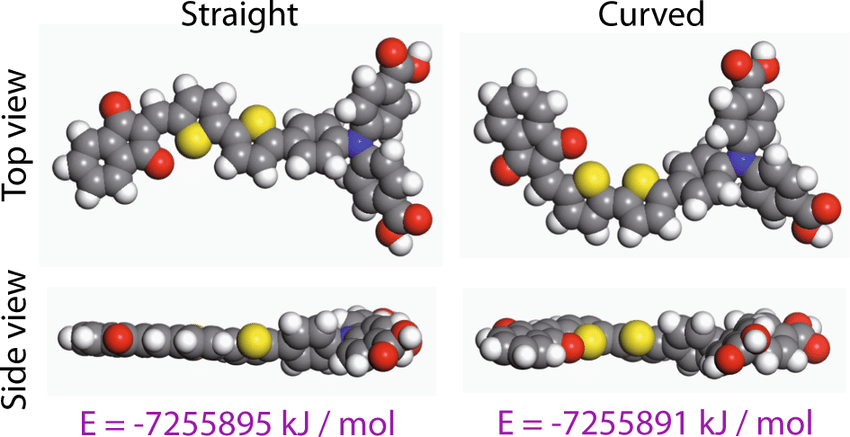}
    \centering
    \caption{The scheme of the molecular conformation and corresponding energy \cite{conformers}.}
    \label{fig:conformers}
\end{figure}

\textbf{Molecular conformer} (presented in \cref{fig:conformers}) is a description of how the atoms in the molecule are oriented in space. Working with conformers allows scientists to include 3D structural information into their work. In many cases, it is essential, because the molecule conformation holds many important features, e.g. enantiomers — two stereoisomers, which are mirror images of each other — rotate the polarized light in the opposite directions, which is important in spectroscopy.

\begin{figure}[h]\includegraphics[width=0.7\textwidth]{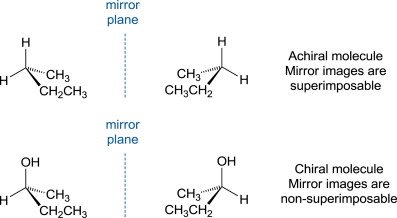}
    \centering
    \caption{Superimposable and non-superimposable molecules \cite{chirality}.}
    \label{fig:chirality}
\end{figure}

\textbf{Chirality} (shown in \cref{fig:chirality}) is a very important property of molecules in stereochemistry. A molecule is chiral if it is distinguishable from its mirror image. In other words, it is not possible to rotate the original molecule in such a way, that it would impose on the mirrored version.

%%%%%%%%%%%%%%%%%%%%%%%%%%%%%%%%%%%%%%%%%%%%%%%%%%%%%%%%%%%%%%%%%%%%%%%%%%%%%%%
\section{Molecular fingerprints examples}
\label{sec:mol-fp-examples}
In this section, we describe the fingerprints that will be implemented in our library. Our choice was based on the popularity and efficiency of those solutions.

%%%%%%%%%%%%%%%%%%%%%%%%%%%%%%%%%%%%%%%%%%%%%%%%%%%%%%%%%%%%%%%%%%%%%%%%%%%%%%%
\newpage
\subsection{Atom Pair}
\label{sec:atom-pair}

\textbf{Atom Pair fingerprint} \cite{atom-pair} is a hashed fingerprint, which counts the atom pairs. Atom pair is defined as a triplet of two (non-hydrogen) atoms and their shortest path distance in a molecular graph. All unique triplets are enumerated and stored in a sparse count simulated or bit vector format.

These features are general enough that a significant number of them can be found in a broad and diverse variety of molecular structures. At the same time, they are specific enough that in the aggregate they can discriminate even closely related topological isomers from one another. For that reason, they can be easily utilized in similarity probing \cite{atom-pair} (search for similar molecules).

\begin{figure}[h]
    \includegraphics[width=0.5\textwidth]{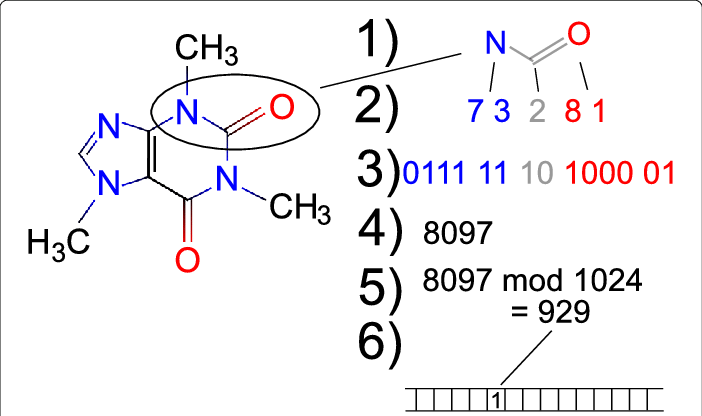}
    \centering
    \caption{The construction of Atom Pair fingerprint \cite{atom-pair}.}
    \label{fig:atom-pair-desc}
\end{figure}

The outline of Atom Pair fingerprint construction is as follows (all these steps can be seen in \cref{fig:atom-pair-desc}):
\begin{enumerate}
    \item Extract all atom pairs
    \item Encode fragments into integers (indexes)
    \item Hash the indexes, using a binary vector of length $n$
    \item For each hashed index turn on the corresponding bit, i.e. bits corresponding to atom pairs present in the molecule are turned on, the remaining bits are turned off
\end{enumerate}

Fingerprint hyperparameters are as follows:
\begin{itemize}
    \item number of bits — by default set to 2048
    \item minimum and maximum distance — the range of distance between 2 substructures to consider them
    \item whether to use count simulation (explained below)
    \item whether to use chirality during description of circular neighborhoods
\end{itemize}

\textbf{Count simulation} \cite{rdkit} is a way of approximating substructure counts, while reducing memory and computational requirements. The idea is to use multiple bits per feature. Each bit represents a certain threshold for counting a substructure. It requires less memory and is faster than count vectors, but at the same time to include counting of the substructures in the molecule.

%%%%%%%%%%%%%%%%%%%%%%%%%%%%%%%%%%%%%%%%%%%%%%%%%%%%%%%%%%%%%%%%%%%%%%%%%%%%%%%
\subsection{ECFP}
\label{sec:ecfp}

\textbf{Extended-connectivity fingerprint (ECFP)} \cite{ecfp} are the one of the most widely used fingerprints.  In particular, their important feature is that they can be rapidly calculated, and at the same time typically give very good results for similarity searching and chemical clustering. This fingerprint inspired many others, e.g. E3FP or MAP4, which are described in the further part of this work. ECFP fingerprint use circular neighborhoods around each atom, similarly to breadth-first search (BFS) algorithm. The radius of this subgraph (or its diameter, depending on the implementation) is an important hyperparameter and is typically denoted when this fingerprint is used. Here, we use the more popular diameter-based notation, so for example by ECFP4 we mean using 2-hop neighborhood (with radius 2, which translates to diameter 4) around each atom. This special case of ECFP4 is the most popular one, and is also called a Morgan fingerprint, because of its relation to Morgan algorithm \cite{morgan-algorithm}. The whole concept of circular fingerprint is primarily inspired by Weisfeiler-Lehman isomorphism test (WL-test) \cite{WL}. ECFP is a typical example of a hashed fingerprint. 

\begin{figure}[h]
    \includegraphics[width=0.7\textwidth]{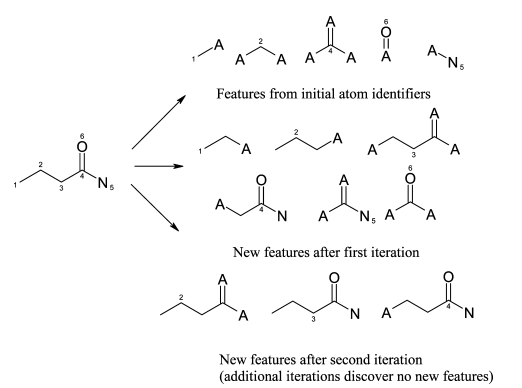}
    \centering
    \caption{Generating new features after consecutive iterations in ECFP fingerprint \cite{ecfp}.}
    \label{fig:ecfp-desc}
\end{figure}

ECFP fingerprint has the following hyperparameters:
\begin{itemize}
    \item number of bits, i.e. the final dimensionality — by  default it is set to 2048
    \item radius, i.e. the circular neighborhood size around each atom taken into consideration
    \item type of resulting vector — bit or count
    \item whether to use chirality during description of circular neighborhoods
    \item whether to consider atom placement inside a ring
\end{itemize}

By default, typically, we use a binary (bit) vector with 2048 bits and radius 2. During hyperparameter tuning, typically the number of bits is the most important, followed by the type of vector. Radius is rarely tuned in practice.

\noindent
\newpage
The outline of the algorithm used to compute this fingerprint is as follows:
\begin{enumerate}
    \item
    Assign a unique identifier to each atom in a molecule, based on a hash function calculated from atom features. Those are, for example:
    \begin{itemize}
        \item number of non-hydrogen immediate neighbors
        \item valency minus the number of connected hydrogens (in other words, total bond order ignoring bonds to hydrogens)
        \item atomic number
        \item atomic mass
        \item atomic charge
        \item number of attached hydrogens (both implicit and explicit)
        \item whether the atom is a part of at least one ring
    \end{itemize}

    \item For each atom identifier in our list, we get all of its neighborhood of a certain radius, concatenate it with the identifier of that atom, and hash all of them to get a new value of the identifier. After each iteration, we append the resulting identifiers to the feature list of our fingerprint. We repeat this step for k-hop neighborhood, i.e. \(k=0, 1, 2, ...\). In \cref{fig:ecfp-desc} we can see substructures obtained during first 2 iterations on an example molecule.
    
    \item Deduplicate the structures, which can arise due to the same structure getting different identifiers in a feature list. It is very likely that the same structures will occur in the feature list multiple times, but under different identifiers. To remove such duplicates, we keep track of bonds that are included in each substructure, so that the resulting feature is unambiguous.

    \item Convert the result to the bit array of a given size, using the modulo operation. We obtain a list of indices, where the resulting bit vector will have 1s. As a result of hashing, there may occur bit collisions. 
\end{enumerate}

To lower the chance of bit collisions occurring, we may want to increase the length of a resulting vector. However, the number of collisions is typically relatively small. Additionally, this can act as an implicit regularization. Therefore, while the higher number of bits preserves more information, it is not always beneficial.

ECFP fingerprint also has a counted version, in addition to the bit (binary) version outlined above. In that case, there are two changes to the algorithm:
\begin{enumerate}
    \item During iterations, we count all the discovered substructures with a dictionary, instead of noting only their existence. The two identifiers, the diameter and also the number of occurrences of that substructure, are concatenated and then hashed.
    \item There is no need for structures deduplication, as duplicates are explicitly counted.
\end{enumerate}

%%%%%%%%%%%%%%%%%%%%%%%%%%%%%%%%%%%%%%%%%%%%%%%%%%%%%%%%%%%%%%%%%%%%%%%%%%%%%%%
\newpage
\subsection{E3FP}
\label{sec:e3fp}

\begin{figure}[H]
    \includegraphics[width=0.8\textwidth]{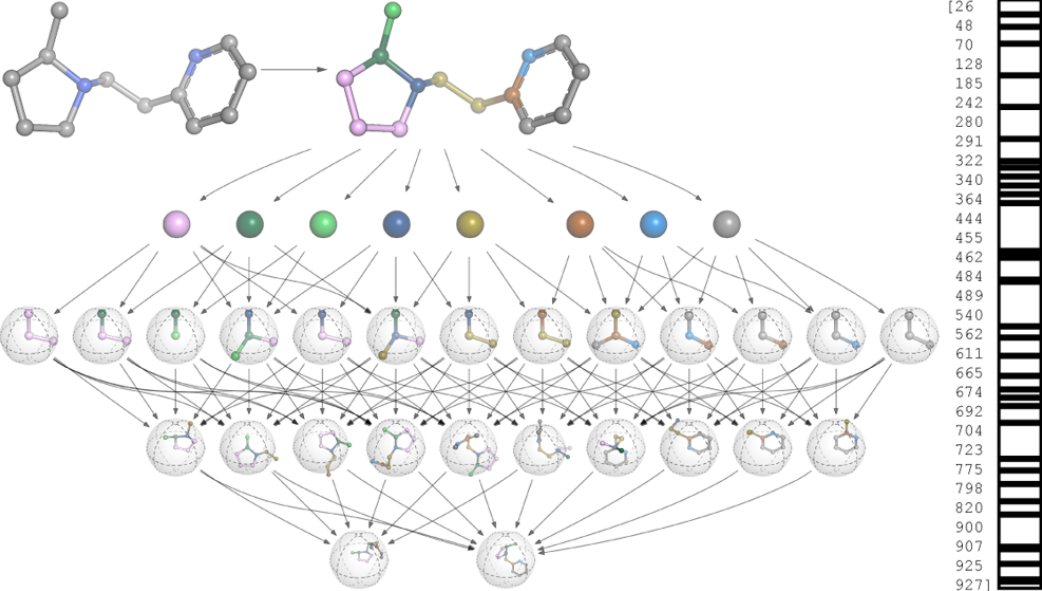}
    \centering
    \caption{The calculation of E3FP fingerprint \cite{e3fp}.}
    \label{fig:e3fp-desc}
\end{figure}

\textbf{Extended three-dimensional fingerprint (E3FP)} \cite{e3fp} is a circular fingerprint designed specifically for modeling 3D molecular structures. The algorithm for its construction is similar to ECFP, but instead of taking a flat 2D atom neighborhood, it uses 3D shells for finding substructures, as shown in \cref{fig:e3fp-desc}.

First, the algorithm generates $k$ conformers with the lowest energy. Low energy signifies a more stable conformer. Many molecules have multiple stable conformations, so it is possible to find several local energy minima. Using these conformers, fingerprints are calculated. We have to aggregate the resulting feature vectors, e.g. take the features based on the conformer with the lowest energy, or average value of each feature for $k$ conformers with the lowest energy. Finally, we can optionally fold the resulting fingerprint.

E3FP is designed for tasks that require taking 3D structure into account like molecule docking prediction, as it includes the structure of conformers. Unfortunately, due to costly computation of three-dimensional conformer generation for E3FP, it is designed to work only for small molecules.

This fingerprint has many hyperparameters, but most of them do not require any tuning, according to the authors \cite{e3fp}. However, few of them may benefit from it:
\begin{itemize}
    \item number of conformers that should be generated
    \item \(k\) - number of conformers with the lowest energy included in the result
    \item maximum difference of energy between conformers — as higher energy indicates less stable conformers, this value prevents the algorithm from including too unstable conformers
\end{itemize}

\clearpage

%%%%%%%%%%%%%%%%%%%%%%%%%%%%%%%%%%%%%%%%%%%%%%%%%%%%%%%%%%%%%%%%%%%%%%%%%%%%%%%
\subsection{ErG}
\label{sec:erg}

\begin{figure}[H]
    \includegraphics[width=0.7\textwidth]{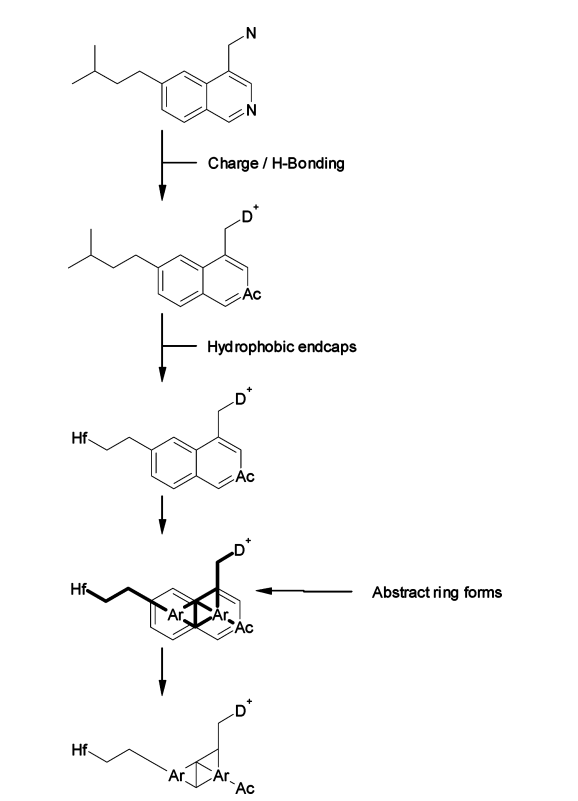}
    \centering
    \caption{The calculation of reduced graph \cite{erg} fingerprint.}
    \label{fig:erg-desc}
\end{figure}

\textbf{Extended-reduced Graph (ErG) fingerprint} \cite{erg} is a 2D pharmacophore description (a part of a molecular structure that is responsible for a particular biological or pharmacological interaction) originally used for scaffold hopping. This fingerprint is based on the idea of reduced graphs, i.e. subgraph of the original molecule containing most crucial features. They are defined by so-called \textbf{Property Points (PPs)}, such as charge, H-bonding, endcap groups (lateral group, that consist of three atoms and have hydrophobic features), or rings. The example of generation of such reduced graph can be seen in \cref{fig:erg-desc}.

\textbf{Scaffold hopping} is the identification of structurally novel compounds starting from known active compounds by modifying the central core structure of the molecule. Fingerprints allow measuring the similarity between molecules, based on various properties, and thus enable an effective search for new compounds.
\newpage
\noindent
The algorithm for this fingerprint is as follows:
\begin{enumerate}
    \item Generate reduced graph — it is a structure containing only the key features of a molecule graph, like hydrophobic endcaps or aromatic rings. Consecutive steps for generation of such structure (that also can be seen in \cref{fig:erg-desc}) are as follows:
    \begin{enumerate}
        \item Compute formal charge for all atoms, in order to represent the molecule under physiological conditions.
        \item Assign \(D+\) atom to H-bond donors (strongly electronegative atom, that is covalently bonded to a hydrogen bond) and \(Ac\) atom to H-bond acceptors (electronegative atom of a neighboring molecule or ion that contains a lone pair that participates in the hydrogen bond)
        \item Add a centroid atom \(Ar\) for each ring, and assign \(Hf\) the atom in place of each hydrophobic endcap group. 
        \item Retain all ring atoms that are substituted, and create
        bonds from each of these atoms to the centroid (the new \(Ar\) atom in the ring).
        \item Retain all bridgehead atoms (that have bonds outside the ring) in the rings, and create bonds from each of these atoms to the centroids.
        \item Remove all nonsubstituted ring atoms, and retain all bonds between the atoms that are retained in steps b) and c).
    \end{enumerate}
    \item Compute descriptor vector from reduced graph. Property Points (PPs) are converted into the form: PP 1 — shortest path in reduced graph — PP 2. This form is very similar to how the atom pairs are encoded in Atom Pair fingerprint.
    \item Fuzzy incrementation — calculate final fingerprint as a count vector of tuples creates in Step 2. During counting, the first neighboring tuples, in terms of distance in the resulting vector, are incremented by \(incr\) (which is a hyperparameter).

The dimensionality of the resulting feature vector is as follows (\(minDist\) and \(maxDist\) is accordingly the minimum and the maximum shortest path between two Property Points in the reduced graph):
\[size(v) = \frac{n(n+1)}{2}(maxDist - minDist + 1)\]

Hyperparameters for this fingerprint are as follows:
\begin{itemize}
    \item \(incr\) — a value by which all the neighboring fields in a vector are incremented during Step 3 of the algorithm
    \item minimum path — minimum shortest path between two PPs to form a feature during Step 3
    \item maximum path — maximum shortest path between two PPs to form a feature during Step 3
\end{itemize}
\end{enumerate}

%%%%%%%%%%%%%%%%%%%%%%%%%%%%%%%%%%%%%%%%%%%%%%%%%%%%%%%%%%%%%%%%%%%%%%%%%%%%%%%
\newpage
\subsection{MACCS Keys}
\label{sec:maccs}

\textbf{MACCS Keys} \cite{mdl-keys} are 166-bit 2D structure fingerprints that are commonly used for measuring molecular similarity \cite{kuwahara-gao-eigenvalue-entropy}. The structures have been created, based on expert knowledge, by MDL Information Systems \cite{mdl-keys}. They are easy to compare, are interpretable for chemists, and they have good predictive power. It can be interpreted as asking certain questions about the characteristic of a molecule, e.g.:
\begin{itemize}
    \item Are there fewer than 3 oxygens?
    \item Is there an S-S bond?
    \item Is there a ring of size 4?
    \item Is there at least one F, Cl, Br or I present?
\end{itemize}

For each question, the answer may be 0 — false or 1 — true. The resulting fingerprint is a bit vector containing answers to those questions. All 166 keys (fragment definitions) are publicly available, e.g. in RDKit \cite{maccs-github}. Notably, this fingerprint does not have any hyperparameters. This is often the case for fingerprints based on molecular descriptors.

%%%%%%%%%%%%%%%%%%%%%%%%%%%%%%%%%%%%%%%%%%%%%%%%%%%%%%%%%%%%%%%%%%%%%%%%%%%%%%%
\newpage
\subsection{MAP4}
\label{sec:map4}

\textbf{MinHashed atom-pair fingerprint up to a diameter of four bonds (MAP4)} \cite{map4} is a circular fingerprint, but using MinHash algorithm for compression, rather than folding, like in ECFP. MinHash allows quick estimation for similarity between two sets. In this case, it is essential, as it enables fast similarity searches in very large databases \cite{map4}. MAP4 also has proven to be better at working with larger molecules in comparison to other fingerprints. 

\begin{figure}[h]
    \includegraphics[width=0.8\textwidth]{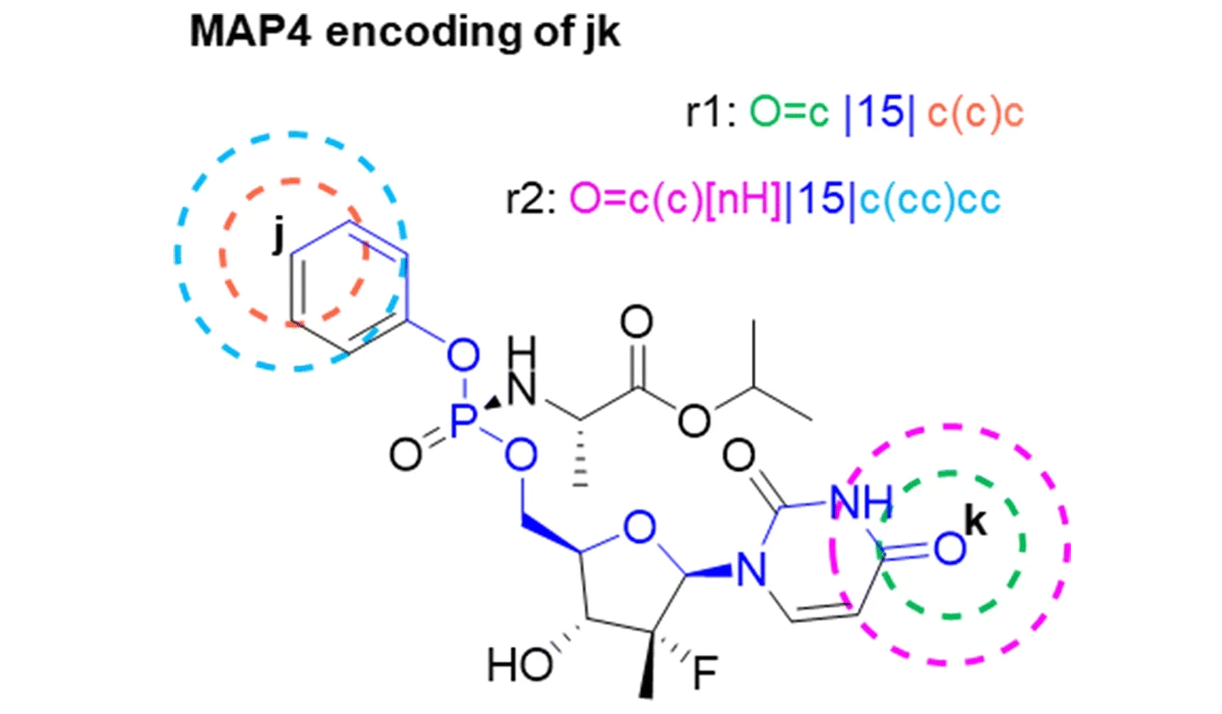}
    \centering
    \caption{MAP4 atom pair encoding \cite{map4}.}
    \label{fig:map4-desc}
\end{figure}

The algorithm for its computation is as follows:
\begin{enumerate}
    \item Finding the circular substructures (i.e. k-hop neighborhood subgraphs) surrounding each non-hydrogen atom in the molecule at radii 1 to \(r\) (which is a hyperparameter).
    \item Calculate the shortest path between each atom pair.
    \item
        Rewrite the atom pairs to the proper form for later hashing:
    
        Define \(CSr(k)\) - as a circular substructure of radius \(r\) around an atom \(k\).
        
        Define \(TP_{j,k}\) - as the shortest path between atoms \(j\) and \(k\).
        
        All atom-pair shingles (i.e. 3-element tuples atom-distance-atom) in a form of \(( CSr(j) | TP_{j,k} | CSr(k) )\) are written for each atom pair and each value of \(r\). \(CSr(j)\) and \(CSr(k)\) are in SMILES format and are placed in a shingle in lexicographical order. We can see that exact process in \cref{fig:map4-desc}.
    \item The resulting set of atom-pair shingles is hashed to a set of integers \(S\) using SHA-1, and its corresponding transposed vector \(S^T\) is finally MinHashed to the form of the MAP4 vector.
\end{enumerate}

Fingerprint hyperparameters are exactly the same as in ECFP.

%%%%%%%%%%%%%%%%%%%%%%%%%%%%%%%%%%%%%%%%%%%%%%%%%%%%%%%%%%%%%%%%%%%%%%%%%%%%%%%
\newpage
\subsection{MHFP}
\label{sec:mhfp}

\begin{figure}[H]
    \includegraphics[width=\textwidth]{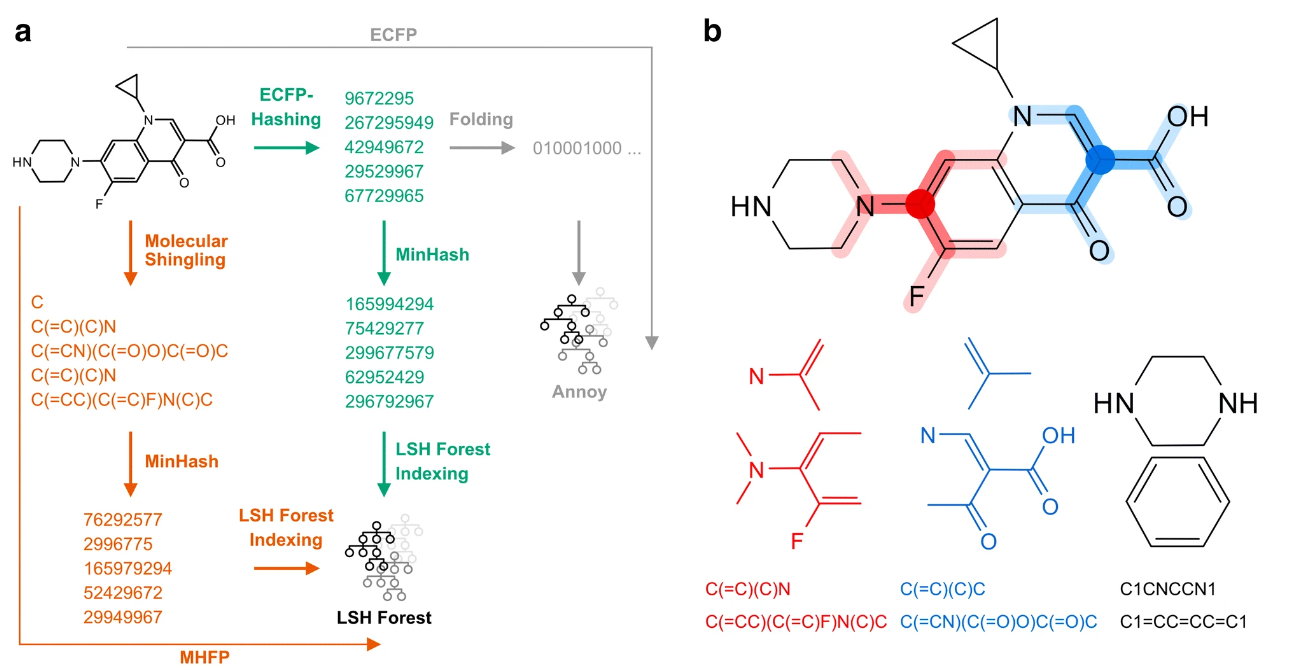}
    \centering
    \caption{MHFP and ECFP workflow comparison \cite{mhfp}.}
    \label{fig:mhfp-desc}
\end{figure}

\textbf{MinHash fingerprint, up to six bonds (MHFP6)} \cite{mhfp} is closely related to MAP4. The algorithms behind them are basically the same — the main difference is that MHFP uses 3-hop neighborhood instead (as can be seen in \cref{fig:mhfp-desc} in subsection b) of 2-hop for MAP4 for finding substructures in a molecule. It is often used with approximate nearest neighbors (ANN) for similarity search.
\clearpage

%%%%%%%%%%%%%%%%%%%%%%%%%%%%%%%%%%%%%%%%%%%%%%%%%%%%%%%%%%%%%%%%%%%%%%%%%%%%%%%
\subsection{Topological Torsion}
\label{sec:topological-torsion}

\textbf{Topological Torsion fingerprint} \cite{nilakantan1987topological} was inspired by the basic conformational element — the torsion angle. This fingerprint aims to capture the predominantly long-range relationships captured in Atom Pair fingerprints by representing short-range information contained in the torsion angles of a molecule. They use four-atom sequences in a form:

\[(NPI-TYPE-NBR) - (NPI-TYPE-NBR) - (NPI-TYPE-NBR) - (NPI-TYPE-NBR),\]

\noindent where \(NPI\) indicates the number of \(\pi\) electrons on each atom, \(TYPE\) indicates the atomic type and \(NBR\) is the number of non-hydrogen branches.

\begin{figure}[H]
    \includegraphics[width=0.8\textwidth]{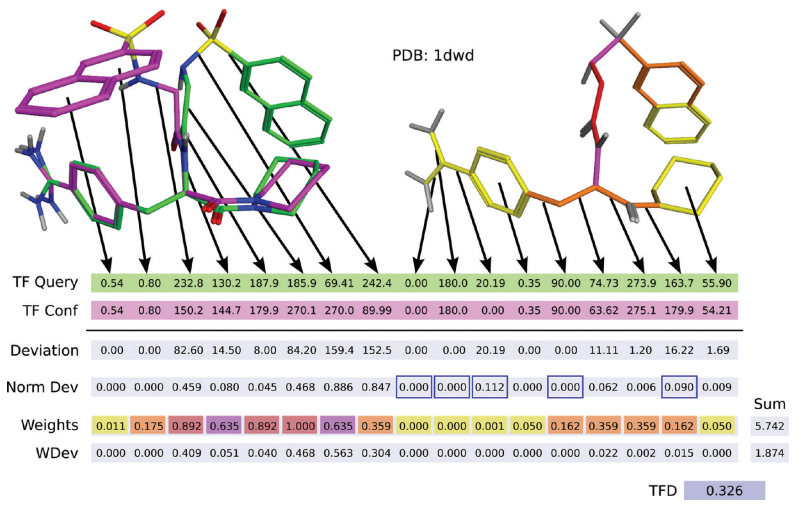}
    \centering
    \caption{Calculation of torsion fingerprint \cite{tfd}.}
    \label{fig:topological-torsion-desc}
\end{figure}

It is a great advantage of Topological Torsion fingerprint, that by using 2D input, it is able to model distant dependencies and include 3D information. We can see in \cref{fig:topological-torsion-desc}, that using this method we are able to estimate the energy function of a conformer for a given molecule — represented by weights on the image.

However, Topological Torsion fingerprint has also some disadvantages — it is not designed to work with large molecules, especially polymers, e.g., proteins, and it can only be accessed in sparse count vector format.

%%%%%%%%%%%%%%%%%%%%%%%%%%%%%%%%%%%%%%%%%%%%%%%%%%%%%%%%%%%%%%%%%%%%%%%%%%%%%%%

\newpage
\section{MoSCoW priorities estimation}
\label{sec:moscow}

\noindent
To create a functionality prioritization for the library, we use the MoSCoW method. In this method, we define 4 levels of priority, each represented by a different letter. Different tasks — in the form of user stories — are then grouped by their assigned priority to allow for easy project planning. The priorities are:
\begin{itemize}
\item M — MUST — the most crucial features needed for the product to be delivered
\item S — SHOULD — important features that would greatly improve products quality
\item C — COULD — features that would be nice to have but are not necessary for project's completion
\item W — WON'T HAVE (THIS TIME) — features that will not be delivered in this project but leave potential for future work
\end{itemize}

\subsection{MUST priority}
\label{sec:must}
\noindent As a chemoinformatician or software engineer, I need:
\begin{itemize}
\item a software to compute multiple commonly used molecular fingerprints
\item to efficiently utilize all cores of my machine and process molecules in parallel
\item all fingerprinting algorithms to be highly configurable, in order to perform hyperparameter tuning and customize them to needs of various projects
\item a Python library compatible with \texttt{scikit-learn} to incorporate in my machine learning workflow
\item to be able to easily download and install library using \texttt{pip} Python package manager
\end{itemize}

\subsection{SHOULD priority}
\label{sec:should}
\noindent As a chemoinformatician or software engineer, I need:
\begin{itemize}
\item a comprehensive documentation so that I can easily understand how to use software library
\item the library to be versioned and available on GitHub so that I can track any changes or report bugs or issues
\item the library to be an Open Source solution which allows me to contribute to it
\item have tools for continuous development and integration of the library so that it is easy to contribute to it and maintain it
\item a library with large suite of unit tests and high test coverage, so that I can be sure that any changes or customizations that I make are working well
\end{itemize}

\subsection{COULD priority}
\label{sec:could}
\noindent As a chemoinformatician or software engineer, I would like to have:
\begin{itemize}
\item readable code with comments, so that I can easily understand underlying logic and customize it to my needs
\item code usage examples to allow me to see how to use the library in my own implementations and solutions
\item a software to compute 3D fingerprints, e.g. E3FP, or 3D pharmacophores
\end{itemize}

\subsection{WON'T HAVE priority}
\label{sec:won't}
\noindent As a chemoinformatician or software engineer, I need:
\begin{itemize}
\item a software implementing fingerprints that take into account a 3D structure of the molecule
\item a software implementing deep-learning algorithms for computation of molecular fingerprints
\item the library to implement other molecule preprocessing methods, such as molecule standardization or transformations between different formats, such as from SMILES to molecular graph.
\end{itemize}

%%%%%%%%%%%%%%%%%%%%%%%%%%%%%%%%%%%%%%%%%%%%%%%%%%%%%%%%%%%%%%%%%%%%%%%%%%%%%%%

\chapter{\ChapterTitleRealizationAspects}
\label{sec:realizacja}

\noindent In this chapter we present the selected realization aspects of our project, such as repository structure and name, package deployment, the ways in which we improve efficiency and tools that allow us to do it, assessment of the code quality and performance as well as DevOps related matters. 

\section{Library}
\label{sec:biblioteka}

\noindent Our library is called \textbf{scikit-fingerprints}. The structure of code repository was planned to be compatible with regular Python library structure. This allows us to automatically integrate it with \texttt{pip} and deploy to PyPI (Python Package Index) package repository.

The code repository consists of two main directories. The \texttt{skfp} directory is the core part of the library which implements all the algorithms. It is the Python package that we import in our Python files or Jupyter Notebooks. Inside it there are \texttt{.py} files containing our implementation of fingerprints as well as all the helper functions and utilities for them.

The testing, performed with files stored in \texttt{tests} directory, asserts that the fingerprints are working correctly. We use \texttt{pytest} tool to implement tests that compare the results of our implementation with the sequentially processed fingerprints.

\section{Fingerprints}
\label{sec:fingerprints}
\noindent The most essential part of our library implementation are the fingerprints. We settled for implementing each of them in a separate class and file to ensure separation of concerns. All of them are fully \texttt{scikit-learn} compatible and implement its \texttt{transformer} interface. In particular, we define \texttt{.fit()} and \texttt{.transform()} methods for each fingerprint. It is worth mentioning that in this case the \texttt{.fit()} method is empty as our transformers are not implementing machine learning algorithms and do not have any trainable parameters. It is implemented only for compatibility with \texttt{scikit-learn} tools. The input to the \texttt{.transform()} method is a list of either SMILES strings or RDKit molecule objects. This method is responsible for calculating fingerprint features for all the inputs, and return them in the same order. Following the required \texttt{scikit-learn} convention, all fingerprint hyperparameters are passed as constructor arguments and are saved as attributes. This allows users to make use of our fingerprints in machine learning pipelines, fully compatible with other preprocessing methods and various ML algorithms. In particular, this makes hyperparameter tuning much easier. We avoid code duplication by using abstract base class (ABC) to implement common functionality, in particular for parallel processing. Inheriting classes for different fingerprints focuses only on computation of their respective features. Computational details are further explained in the \cref{sec:efficiency}.

ECFP, Atom Pair and Topological Torsion fingerprint, inheriting from base class, implement a special \texttt{.\_get\_generator()} method, returning a function object called \texttt{generator} to calculate a fingerprint. Generators implement a uniform interface for multiple fingerprints and improve efficiency. In particular, using them allows us to easily use object-oriented programming and functional programming abstractions to avoid code duplication. Generators are implemented with RDKit. Our library is tightly integrated with it, as it is the most commonly used tool for chemoinformatics in Python. The functionalities covered by it apply to a much wider variety of problems than just molecular fingerprints. Because of that, we expect users to also use functionalities from RDKit not covered by our \textbf{scikit-fingerprints} library.

For computation of fingerprints from RDKit that do not implement generators, namely MACCS Keys and ErG fingerprint, we use direct methods. It should be noted that this disparity inside RDKit is exactly one of the reasons for existence of our \textbf{scikit-fingerprints} library. We offer unified interface for all types of molecular fingerprints. Additionally, we also offer some fingerprints that are not implemented in RDKit. In our library we reimplement MAP4 and MHFP fingerprints, in particular to avoid many needless dependencies from the original \texttt{map4} library. It was implemented by researchers who created a paper describing it, hence it is not of particularly good quality as a standalone library.

%%%%%%%%%%%%%%%%%%%%%%%%%%%%%%%%%%%%%%%%%%%%%%%%%%%%%%%%%%%%%%%%%%%%%%%%%%%%%%%

\section{Computational Efficiency Solutions}
\label{sec:efficiency}
\noindent We implemented multiple efficiency optimizations such as parallelism, code optimizations and use of Python-related improvements.

The problem of optimizing the computation time of large datasets with popular molecular fingerprint algorithms can be described as embarrassingly parallel \cite{embarrassingly-parallel}. This means that the obvious solution to this problem is the use of data parallelism. Instead of processing the dataset sequentially, we can split it into parts of equal or almost equal numbers of molecules and perform computation in parallel, potentially achieving a great time performance improvement. Python creates an obstacle for implementing this kind of solution in the form of its Global Interpreter Lock (GIL). It enforces the multiprocessing approach to parallelism as opposed to simply using multithreading. Fortunately, the popular Joblib library \cite{joblib} for lightweight parallel processing in Python allows us to do this very efficiently, as described below. 

The processing of the dataset is handled in the \texttt{.transform()} method of the base \texttt{FingerprintTransformer} class that other fingerprints inherit from. The data that gets passed to the method is split into a number of parts of equal or almost equal size. For each of those parts, using the Joblib library, we manage (i.e. spawn, monitor, and kill) workers. In each process, we compute fingerprints with \texttt{.\_calculate\_fingerprint()} method that each of the fingerprint classes implement. Each worker process calculates fingerprints in its batch sequentially. Joblib can automatically detect the number of physical and logical CPU cores. This ensures optimal usage of the machine's resources, in particular full utilization of all its cores. Additionally, it uses direct memory mapping of the processed data to pass it between processes. This is possible for NumPy arrays because of their C implementation.  In Python, default \texttt{list} objects are implemented as pointers to scattered parts of memory, which allows for flexible resizing, but it is also suboptimal for many computationally-heavy tasks such as our solution. In contrast, in C programming language, memory space allocated for one array or matrix is a contiguous part of machine memory. Because of this, Joblib can efficiently pass data between processes.

Moreover, our algorithm aggregates the results of computation only once.  We perform this synchronization in the body of \texttt{.transform()} method after all workers finish their work, as monitored by Joblib. These features let us avoid unnecessary memory reallocation and excessive copy operations, thus improving the performance even further.

Python as the programming language has numerous ways of optimizing the code. These methods are both native to the language or can be used with third-party packages. We do our best to use these tools as much as possible to optimize the efficiency of our solution. Firstly, we try to follow the best code optimization and efficiency practices. Our library breaks out of loops whenever it is possible. It also uses list comprehension — a Python feature that in general is faster and easier to read than a typical for loop, as they create the result list without extending the previously initialized Python list object with unnecessary calls to its \texttt{.append()} method. This optimization is especially noticeable with larger datasets. We also use solutions not native to the Python language, such as NumPy arrays. 
As opposed to Python list objects, they are implemented in C language and use numerous low-level optimizations such as optimized BLAS operations for linear algebra. It makes NumPy a perfect tool for mathematical computations and allows our library to perform better.

%%%%%%%%%%%%%%%%%%%%%%%%%%%%%%%%%%%%%%%%%%%%%%%%%%%%%%%%%%%%%%%%%%%%%%%%%%%%%%%

\section{Quality assurance}
\label{sec:testing}

\noindent In our tests, we use a data-driven approach to testing, using real-world benchmark data to make sure our algorithms give correct results. In particular, we make sure that our parallelized implementations exactly match the results given by sequentially computed fingerprints. For this purpose, we use the HIV dataset from MoleculeNet benchmark \cite{moleculenet}. The test is performed for each fingerprint, and if the fingerprint has counted and bit vector output type, we test both of them. This way, we maintain high test coverage. We perform the computation with fingerprint transformers and sequential methods for the entire HIV benchmark dataset. After that, we compare them element-wise to check if all elements are the same. To implement the tests, we used \texttt{pytest} framework.

To ensure high code quality during development, we use pre-commit hooks. The pre-commit is a tool that runs scripts and checks every time a Git commit is made in a given repository. The hooks that we use are:
\begin{itemize}
    \item black \cite{black} — code formatter that ensures the code style is uniform within the project
    \item isort \cite{isort} — code formatting tool specifically for sorting imported packages in compliance with PEP8 code standard \cite{pep8}
    \item Xenon \cite{xenon} — measures cyclomatic (McCabe) complexity score of our code and halts the commit if the complexity is too high
    \item Bandit \cite{bandit} — security linter for finding common security issues in Python code, which prevents developers from accidentally committing unsafe code
\end{itemize}

%%%%%%%%%%%%%%%%%%%%%%%%%%%%%%%%%%%%%%%%%%%%%%%%%%%%%%%%%%%%%%%%%%%%%%%%%%%%%%%
\newpage
\section{DevOps and package deployment}
\label{sec:devops}

\noindent One of the fundamental aspects of our project revolved around product distribution and accessibility to a wider audience, potentially increasing its visibility within the Python community, therefore simplifying integration into other projects and streamlining research efforts in the field of chemoinformatics. The core features of our solution were designed to ensure a seamless installation process and user-friendly operation, simplifying its use and aligning it with the experience of utilizing other Python packages. In particular, this includes dependency management, library publishing via CI/CD process, and easy installation by the end user. We greatly simplified this process by using Poetry tool.

%%%%%%%%%%%%%%%%%%%%%%%%%%%%%%%%%%%%%%%%%%%%%%%%%%%%%%%%%%%%%%%%%%%%%%%%%%%%%%%

\subsection{Poetry}
\label{sec:poetry}

\noindent Poetry is a Python dependency management and packaging tool. It enabled us to create, build, and publish our project while automatically handling the required dependencies.

Poetry project is configured via \texttt{pyproject.toml} file in a simple, human-readable TOML format. It contains the dependencies list, including optional constraints on their versions, and is read by Poetry in order to install and manage all required dependencies when the library is downloaded. Dependencies in this file are divided into various groups based on their utility, such as main and development group dependencies, enabling separation of dependencies for different purposes. This is very useful for management of the project. This file additionally contains the configuration for the library's structure, its name and authors. Poetry upon issuing \texttt{poetry lock} command resolves dependencies, i.e. it builds the graph of immediate and transitive dependencies, taking into consideration provided constraints. It automatically checks for newest versions satisfying requirements, and saves the complete results to \texttt{poetry.lock} file. It contains the exact version of each dependency for our library. This file ensures the consistency of project environment between developers, and reduces errors from versions mismatch.

During the development of our library, the standard project configuration, when integrated with this tool, operates as follows: we utilize the \texttt{poetry install} command to install all project dependencies and make modifications as needed. Once the dependencies are installed, we can use the \texttt{poetry build} command to create source and wheel packages. The \texttt{poetry.lock} file is created and updated during the installation process to reflect the exact versions of the dependencies used. The process of building the project and managing the \texttt{poetry.lock} file is distinct from the process of publishing the package, which is accomplished using the \texttt{poetry publish} command. This distinction is important as it separates the workflow into two parts: typical development work and package publication.

%%%%%%%%%%%%%%%%%%%%%%%%%%%%%%%%%%%%%%%%%%%%%%%%%%%%%%%%%%%%%%%%%%%%%%%%%%%%%%%

\subsection{Continuous Integration/Continuous Delivery (CI/CD)}
\label{sec:ci-cd}

To achieve an efficient and time-saving project development process, we employed automation through CI/CD tools. Such tools automate all repeated steps in the development project, which in our case were testing and releasing next versions of the project. Specifically, we used GitHub Actions. It allows defining and executing various scripts in response to events within the repository. This way, we could automate the processes of building, testing, and deploying software.

We implemented automated testing procedures and the publication of the library on the PyPI platform upon the release of a new library version. When a new code is added to the development branch, a script is triggered, automatically running tests on the library. If the tests are successful, the developer can safely merge all new changes to the main branch. When changes are released, another script uploads the library to the TestPyPI environment, where all the product release and installation processes can be verified. When all checks are completed, the developers have the option to release a new library version by adding a tag to the new product release and executing the scripts integrated into our library. To label the changes, we use semantic versioning \cite{semver}, a widely-adopted version scheme that encodes a version by a three-part version number MAJOR.MINOR.PATCH, an optional pre-release tag, and an optional build meta tag. In this scheme, risk and functionality of new functionalities are the measures of significance. The first release was tagged as a 1.0.0 release.

%%%%%%%%%%%%%%%%%%%%%%%%%%%%%%%%%%%%%%%%%%%%%%%%%%%%%%%%%%%%%%%%%%%%%%%%%%%%%%%

\subsection{Project licensing and regulations}
\noindent Given that the library is widely accessible \cite{skfp}, allowing anyone to contribute to its development, a series of formal principles were also applied to guide its future growth. We adopted the MIT license \cite{mit-license}, precisely outlined in \texttt{LICENSE.md}, to regulate the legal aspects of third-party use and development of the product. The MIT license (Massachusetts Institute of Technology) is an open and permissive software license, known for its flexibility and ease of comprehension. It permits unrestricted copying, modification, publication, and distribution of our source code, whether in open source or commercial projects, as long as the copyright notice and license information are retained.

Furthermore, we established guidelines promoting a high standard of conduct for community participation in the project. \texttt{CODE\_OF\_CONDUCT.md} outlines permissible and unacceptable behaviors in collaborative work on the product.

We also provided a \texttt{CONTRIBUTING.md} instruction document with precise guidelines on how to propose changes to the product for further improvement. This ensures correct order in development and mitigates potential errors stemming from third-party code changes.

The library also maintains a changelog in \texttt{CHANGELOG.md}, documenting changes in each of its releases. This facilitates the flow of information between developers and users, keeping them informed about newly added features.

%%%%%%%%%%%%%%%%%%%%%%%%%%%%%%%%%%%%%%%%%%%%%%%%%%%%%%%%%%%%%%%%%%%%%%%%%%%%%%%
\chapter{\ChapterTitleResults}
\label{sec:wyniki}

\noindent In this chapter, we present the outcomes of our project. We describe the implemented library, its qualities, performed benchmarking and code quality checks.

\section{scikit-fingerprints library}
\label{sec:product}
\noindent We delivered a software library in Python language, fully compatible with the language ecosystem. We called it \texttt{scikit-fingerprints} to follow the naming convention used by many libraries that focus on compatibility with \texttt{scikit-learn}. To enable an easy installation, we use Python Package Index (PyPI). It serves as the central repository for Python programming language software. It provides a platform where developers can publish and distribute packages, making them easily installable and usable by other developers. To install the most recent version of the library as a PyPI package and verify and install all required dependencies (including transitive dependencies), the only required command is:
\begin{verbatim}
pip install scikit-fingerprints
\end{verbatim}

Then, ready to use, it can be easily imported into the user's Python environment as follows:
\begin{verbatim}
import skfp
\end{verbatim}

Imported package is called skfp. This is because Python package names cannot contain hyphens. The abbreviated name has been inspired by scikit-learn, which is imported as \texttt{sklearn}.

Using the fingerprint transformer was highly streamlined. In order to calculate fingerprints, a user must first create an instance of a class object corresponding to their desired fingerprint. All the fingerprints are accessible from the imported skfp package. Their hyperparameters can be passed as constructor arguments. For example, a boolean \texttt{sparse} parameter determines whether the desired output should be returned as a sparse or dense array object.
\begin{verbatim}
ecfp_transformer = skfp.ECFP(radius=3, sparse=True)
\end{verbatim}
To transform the molecules, the user can pass either a list of molecule objects or SMILES strings to the \texttt{transform} method of the created object. This method returns either NumPy or SciPy's sparse array object of calculated fingerprints, depending on constructor arguments. These fingerprints can then be used further in user's machine learning workflow. As opposed to typical transformer objects used with scikit-learn library, our fingerprint transformers do not require the user to call the \texttt{.fit()} method. However, it is still available in order to provide compatibility with scikit-learn interface. For example, this allows our fingerprints to be a part of scikit-learn's pipeline.
\begin{verbatim}
X_transformed = ecfp_transformer.transform(X)
\end{verbatim}

\section{Code quality maintenance}
\label{sec:quality-and-maintenance}

\noindent We use various tools dedicated for Python development that helped us elevate the quality of our code. The tools we used allowed us to format the code to follow the PEP 8 style guidelines \cite{pep8}, measure the code complexity and ensure safety and security of our code. Black \cite{black} and isort \cite{isort} tools are included as pre-commit hooks to format the code, improving the readability and making the library easier to develop and maintain. We reached a low cyclomatic (McCabe) complexity of 2.27, as measured by Radon \cite{radon} tool for Python. Additionally, Radon offers a descriptive grade of complexity, and our code gets mark “A” for low complexity. This means that code should be relatively easy to understand and maintain.

On top of that, we use a Bandit \cite{bandit} tool to ensure our code is safe and free of many common security issues. The chemical industry is full of patents and regulations, so open-source solutions are not a common thing. To build trust with potential users and ensure our library can be used safely by chemoinformaticians, the security audit is necessary.

Another tool integrated into the project is Flake8 \cite{Flake8}. Flake8 is a static code analysis tool in the Python language. Its purpose is to automatically verify compliance with selected coding standards, such as PEP 8, which defines Python code formatting rules. Flake8 can also detect and signal potential errors and inconsistencies in the code, assisting us in maintaining a high level of source code quality.

\section{Computational efficiency benchmarking}
\label{sec:time-benchmark}
\noindent We tested the time performance of our implementation on the benchmark dataset. We used the HIV dataset \cite{moleculenet-hiv} from MoleculeNet benchmark \cite{moleculenet} that contains 41127 molecules in SMILES string format. In particular, this dataset is quite large by molecular property prediction standards, which allows us to measure scalability of our algorithms. The dataset properties have were summarized in Table \cref{table:dataset-properties}.

\begin{table}[H]
\begin{center}
\begin{tabular}{|c c c c c|} 
 \hline
 Dataset & \# Graphs & Avg. \# Nodes & Avg. \# Edges & \# Classes \\ [0.5ex] 
 \hline\hline
 HIV & 41127 & 25.5 & 27.5 & 2 \\
 \hline
\end{tabular}
\caption{The properties of HIV dastaset from MoleculeNet benchmark.}
\label{table:dataset-properties}
\end{center}
\end{table}

\begin{figure}[H]
  \centering
  \includegraphics[width=0.9\textwidth]{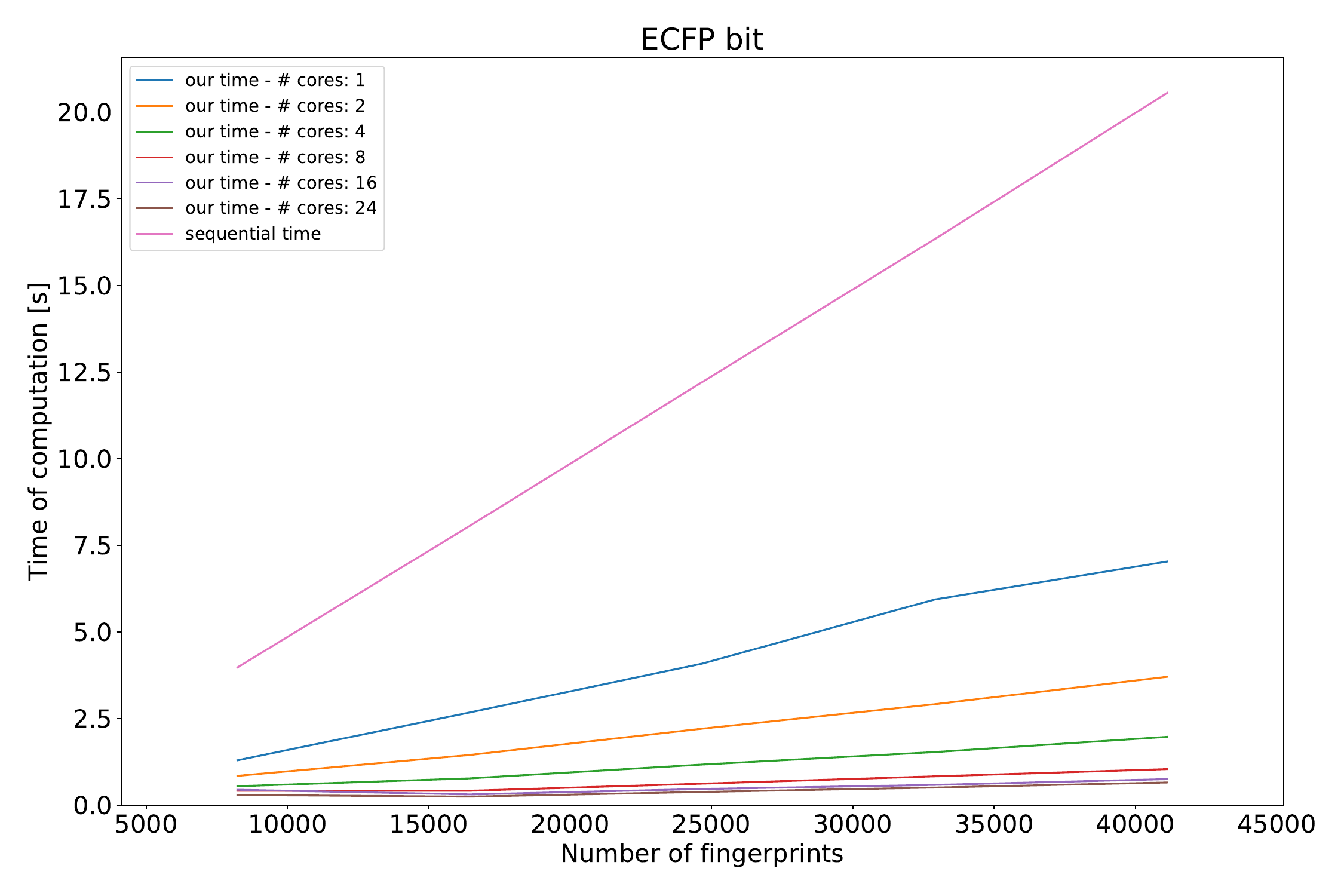}
  \centering
  \caption{Time results for ECFP Fingerprint — bit vector variant.}
  \label{fig:morgan-bit}
\end{figure}

\begin{figure}[H]
  \centering
  \includegraphics[width=0.9\textwidth]{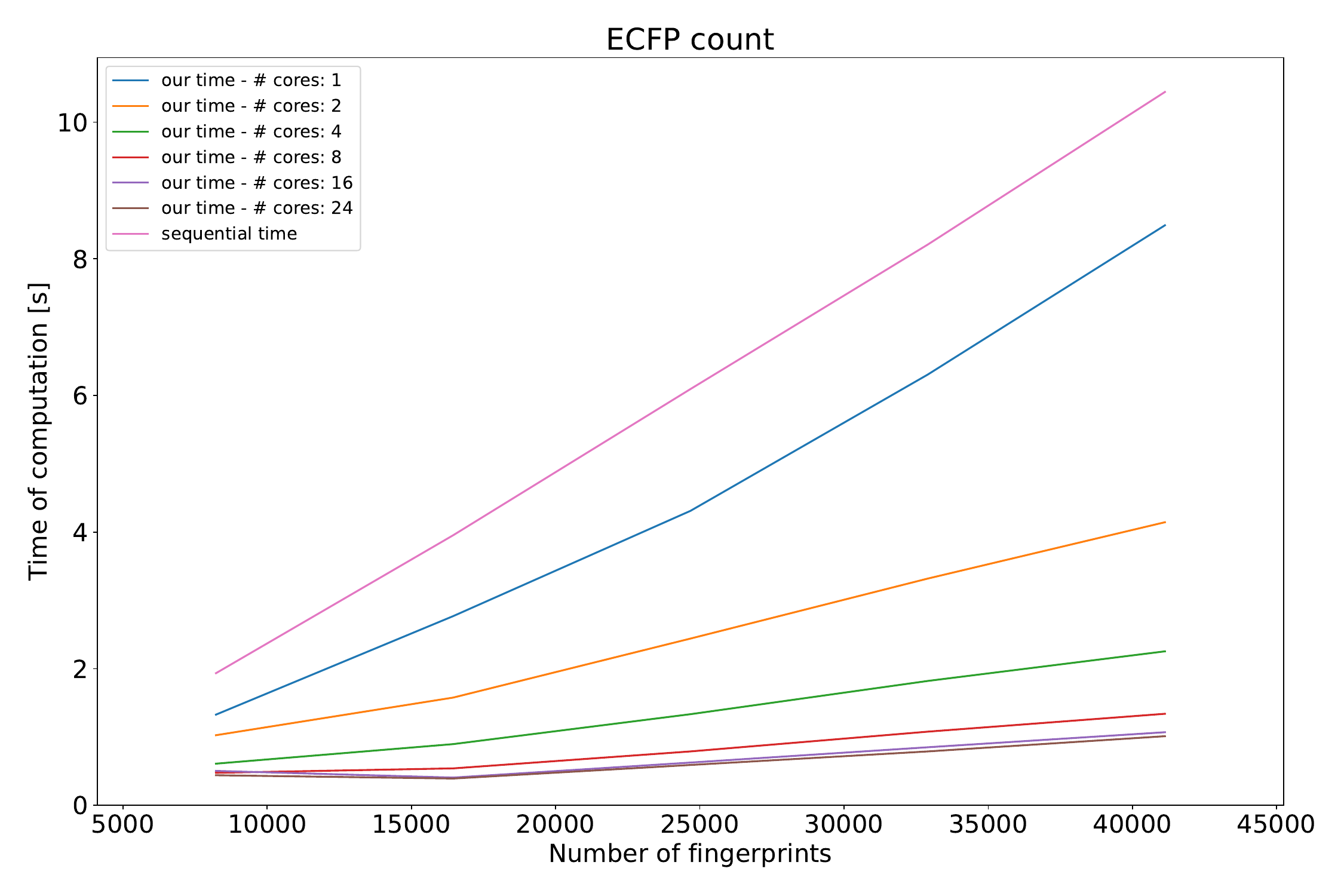}
  \centering
  \caption{Time results for ECFP Fingerprint — count vector variant.}
  \label{fig:morgan-count}
\end{figure}

\begin{figure}[H]
  \centering
  \includegraphics[width=0.9\textwidth]{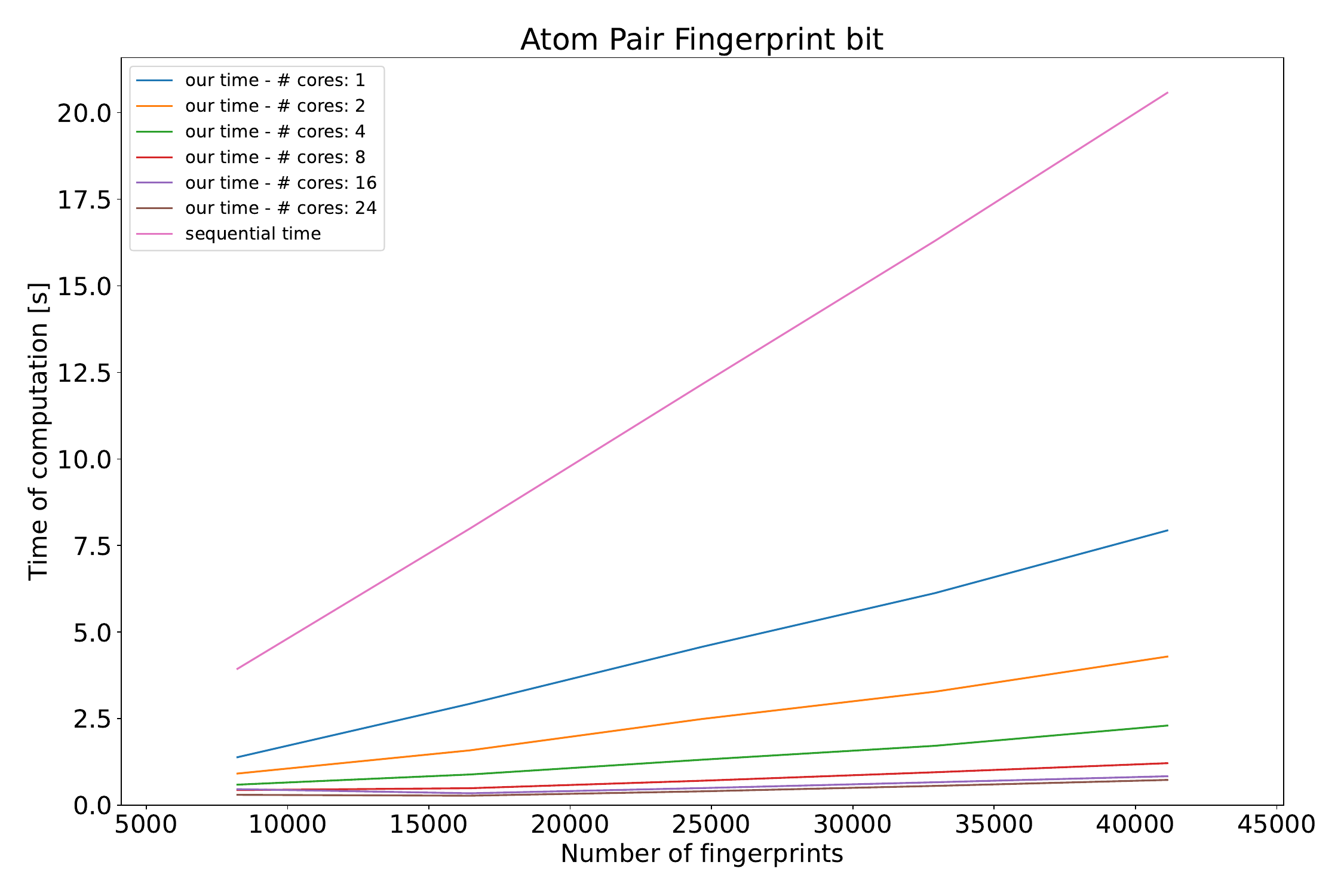}
  \centering
  \caption{Time results for Atom Pair fingerprint — bit vector variant.}
  \label{fig:atom-pair-bit}
\end{figure}

\begin{figure}[H]
  \centering
  \includegraphics[width=0.9\textwidth]{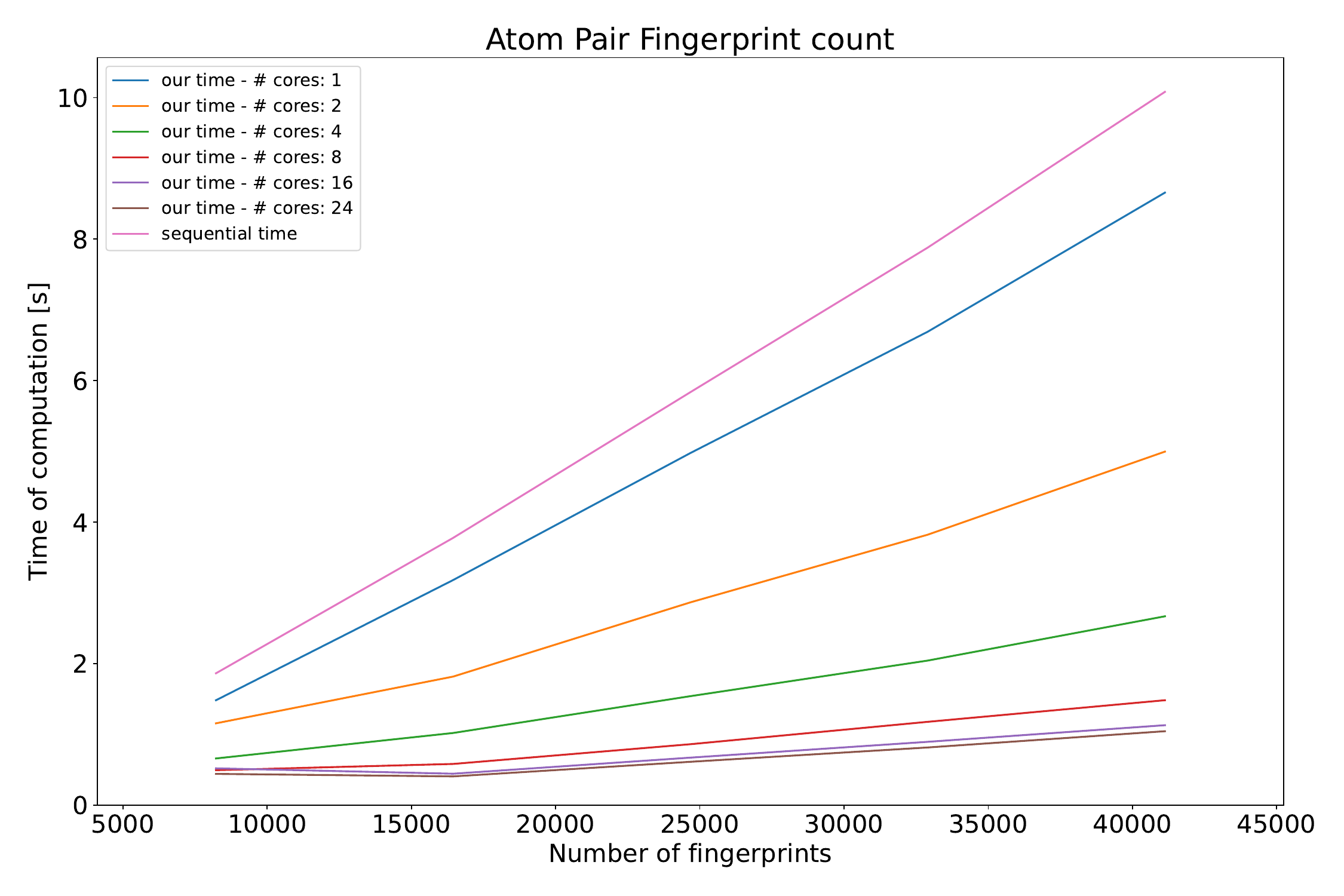}
  \centering
  \caption{Time results for Atom Pair fingerprint — count vector variant.}
  \label{fig:atom-pair-count}
\end{figure}

\begin{figure}[H]
  \centering
  \includegraphics[width=0.9\textwidth]{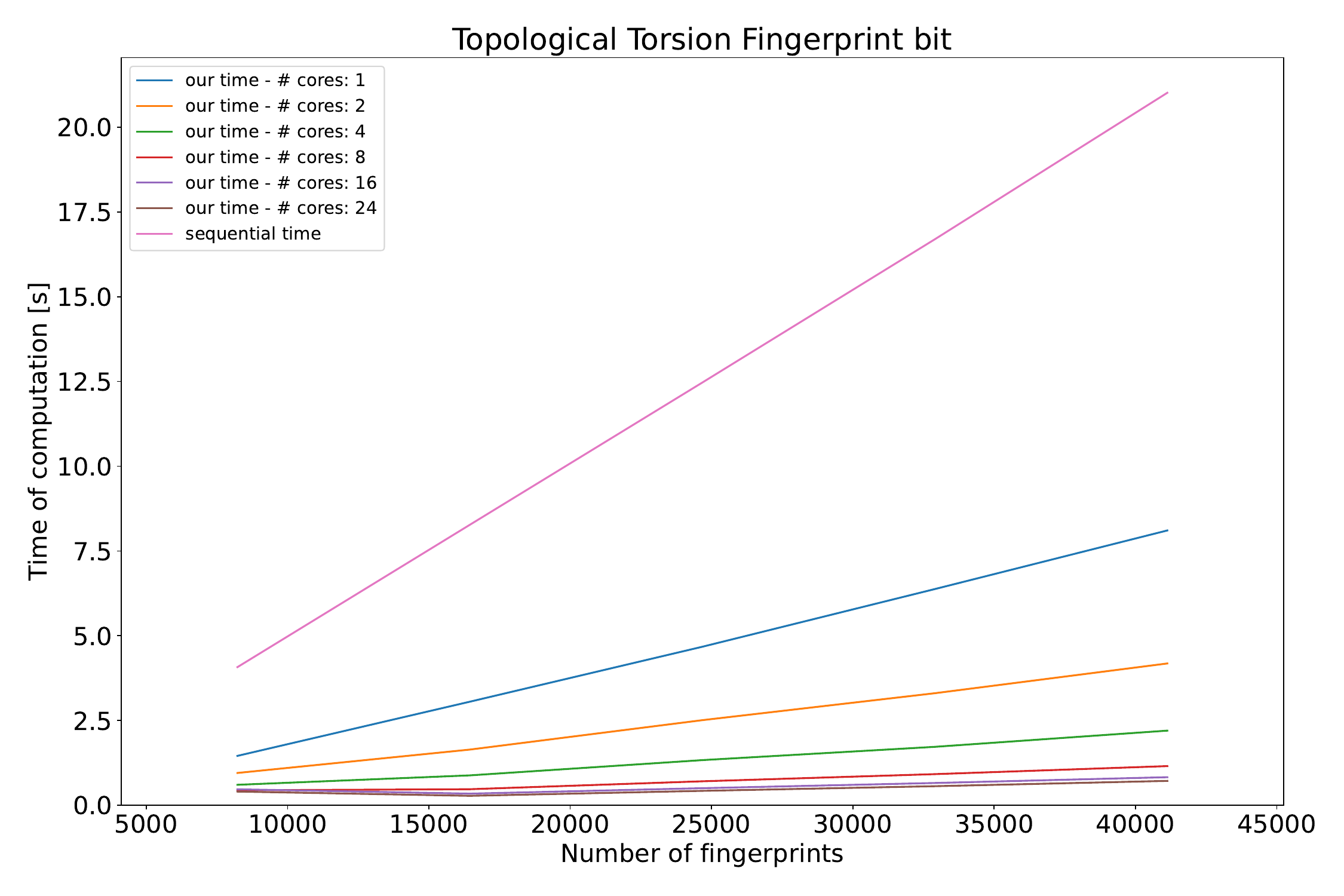}
  \centering
  \caption{Time results for Topological Torsion Fingerprint — bit vector variant.}
  \label{fig:topological-torsion-bit}
\end{figure}

\begin{figure}[H]
  \centering
  \includegraphics[width=0.9\textwidth]{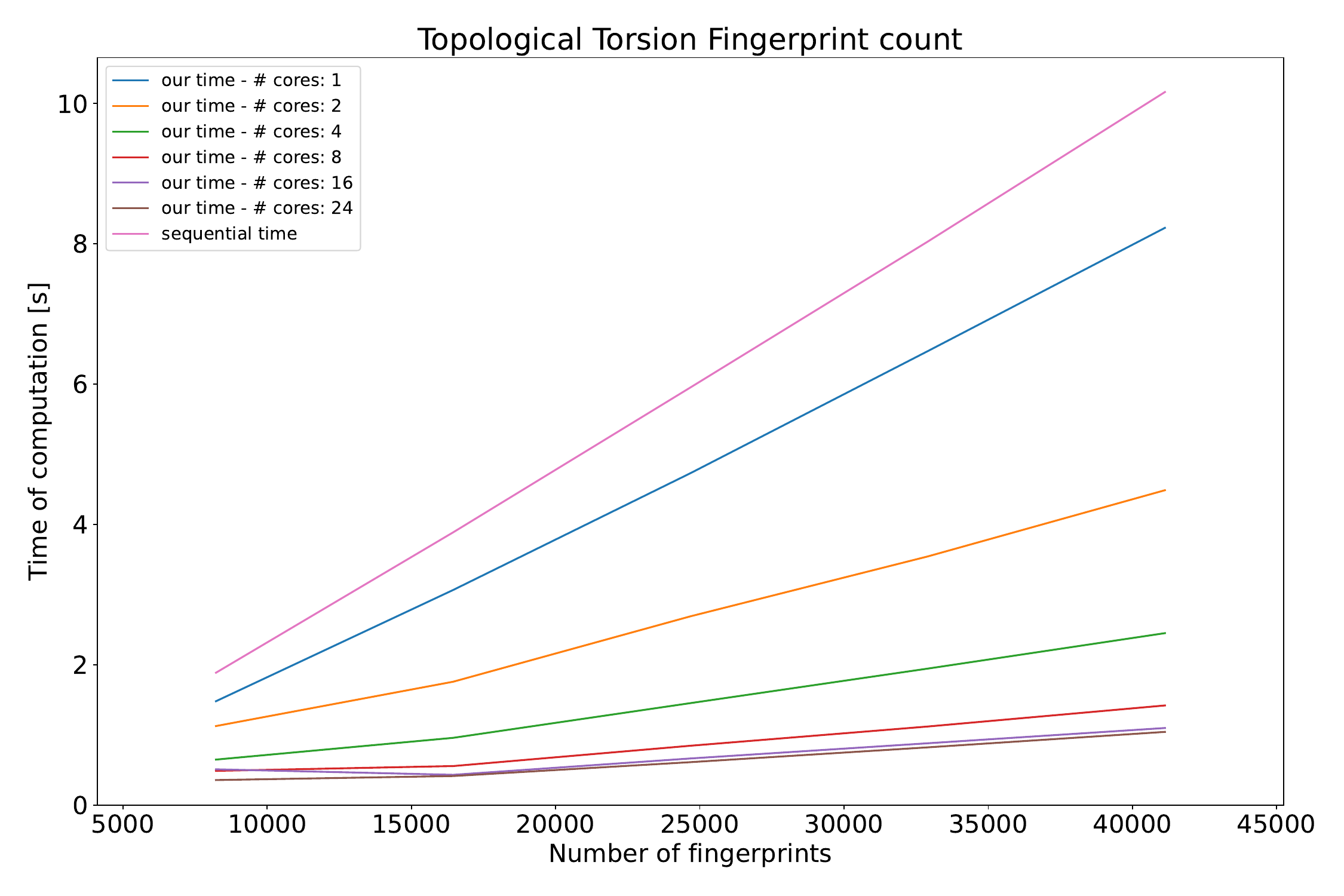}
  \centering
  \caption{Time results for Topological Torsion — count vector variant.}
  \label{fig:topological-torsion-count}
\end{figure}

\begin{figure}[H]
  \centering
  \includegraphics[width=0.9\textwidth]{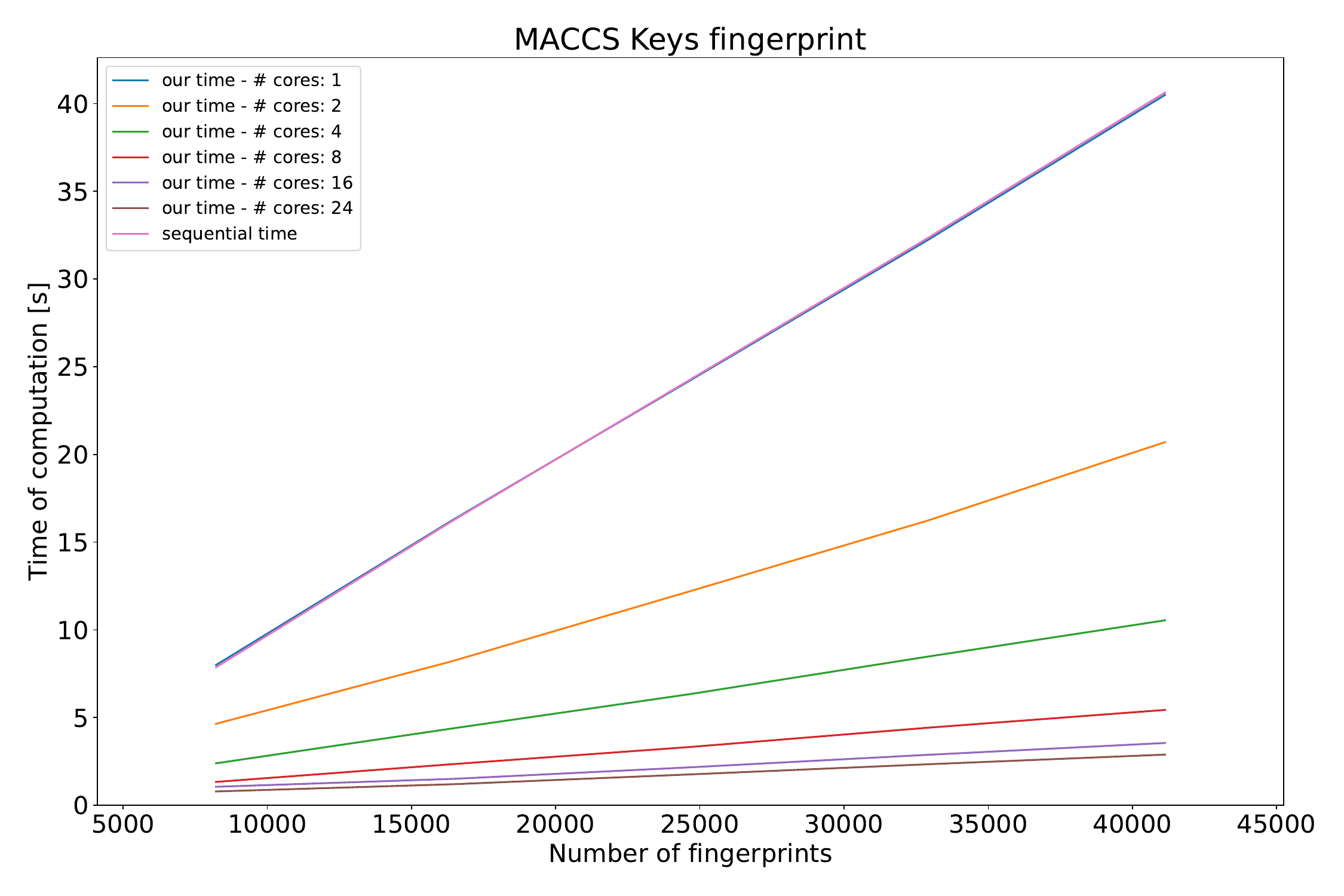}
  \centering
  \caption{Time results for MACCS Keys Fingerprint.}
  \label{fig:maccs}
\end{figure}

\begin{figure}[H]
  \centering
  \includegraphics[width=0.9\textwidth]{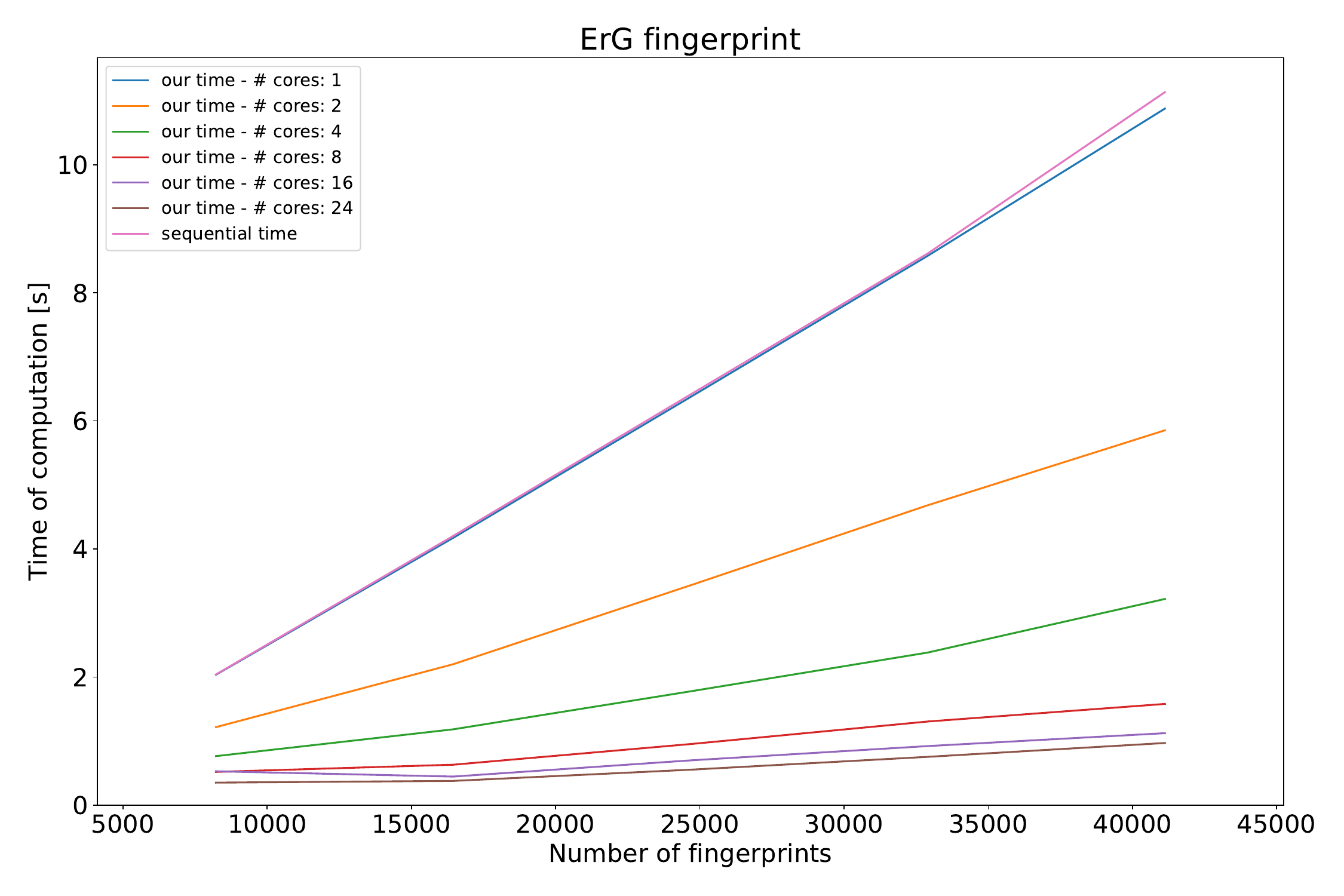}
  \centering
  \caption{ Time results for ErG Fingerprint.}
  \label{fig:erg}
\end{figure}

\begin{figure}[H]
  \centering
  \includegraphics[width=0.9\textwidth]{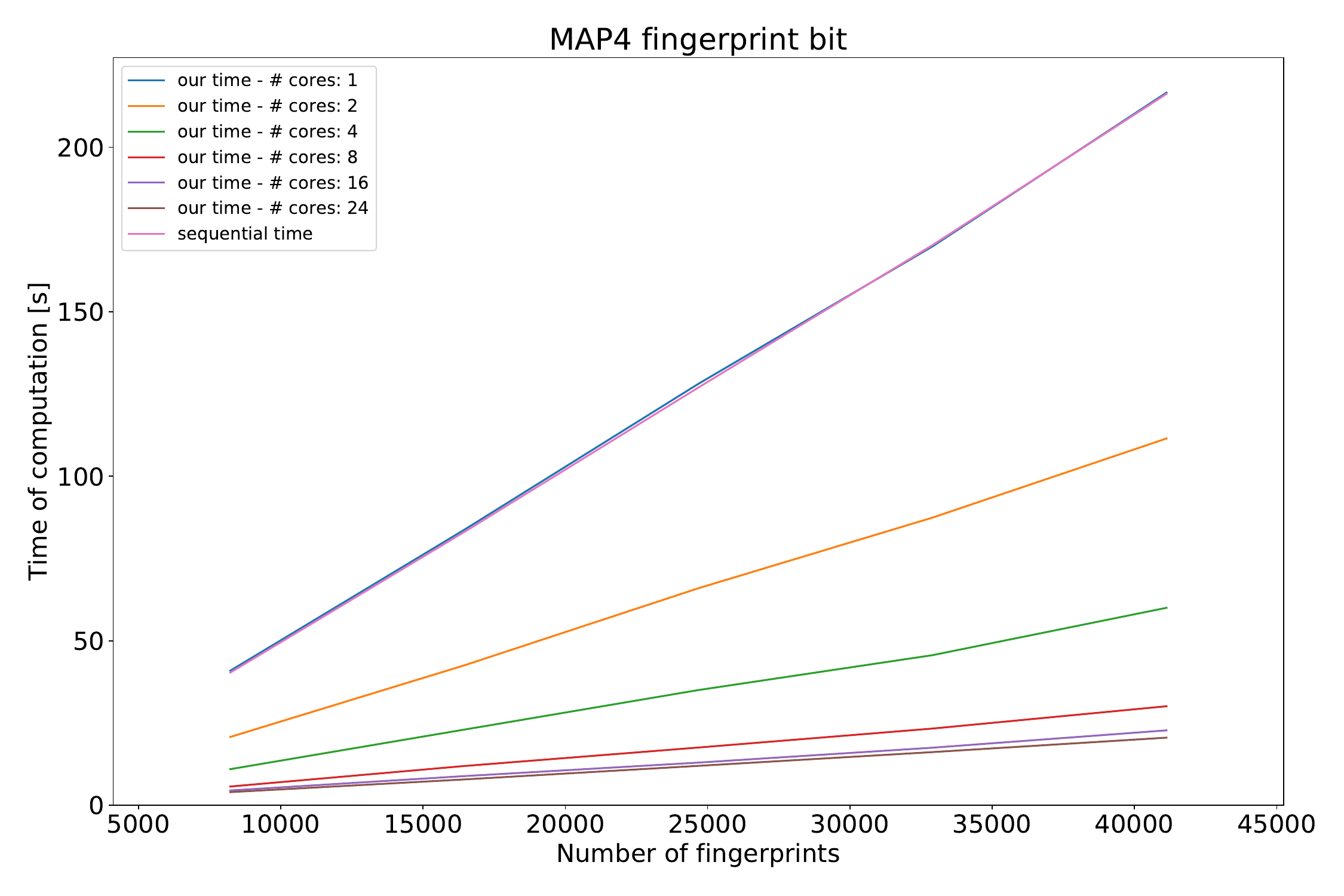}
  \centering
  \caption{Time results for MAP4 fingerprint — bit vector variant.}
  \label{fig:map4-bit}
\end{figure}

\begin{figure}[H]
  \centering
  \includegraphics[width=0.9\textwidth]{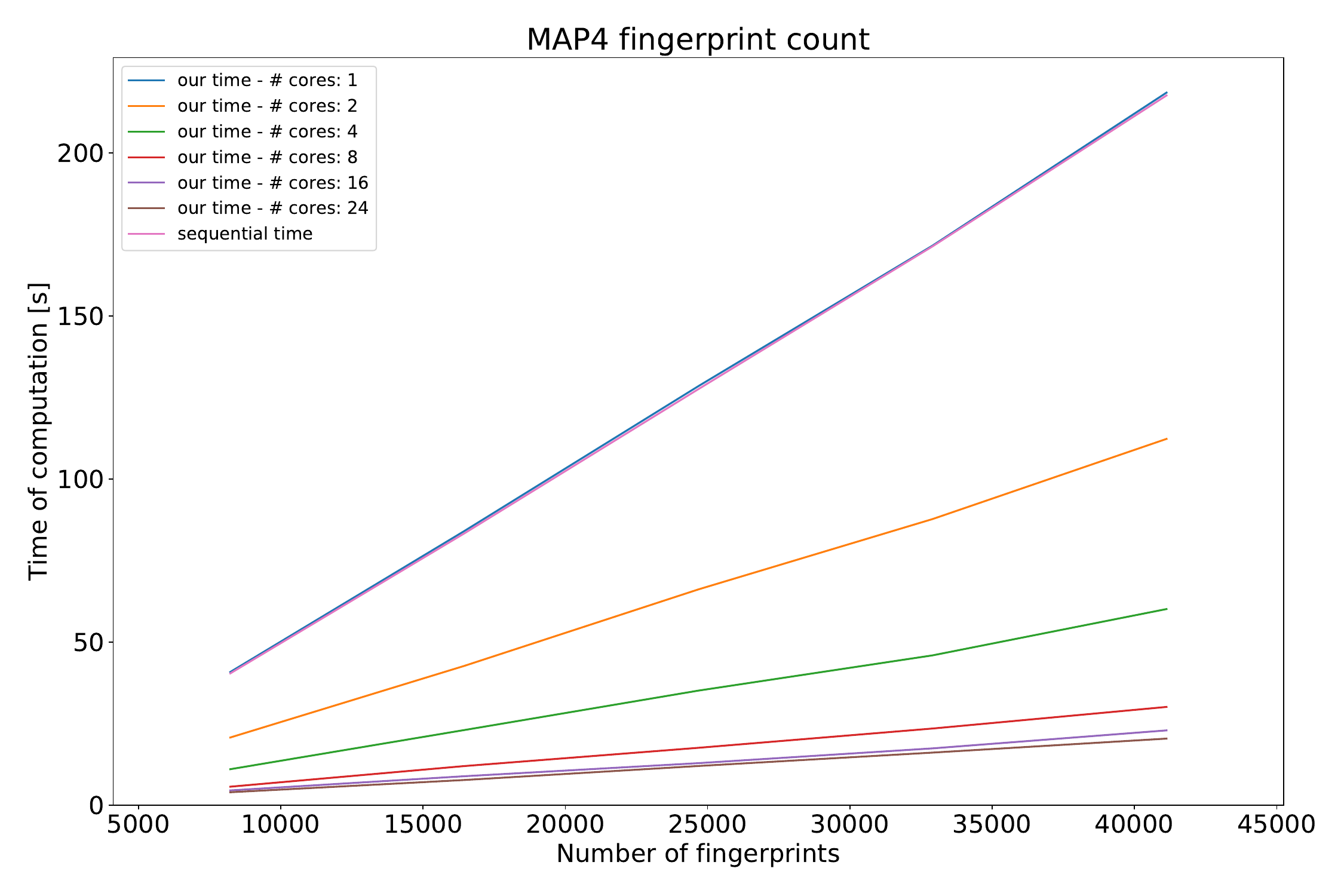}
  \centering
  \caption{Time results for MAP4 fingerprint — count vector variant.}
  \label{fig:map4-count}
\end{figure}

\begin{figure}[H]
  \centering
  \includegraphics[width=0.9\textwidth]{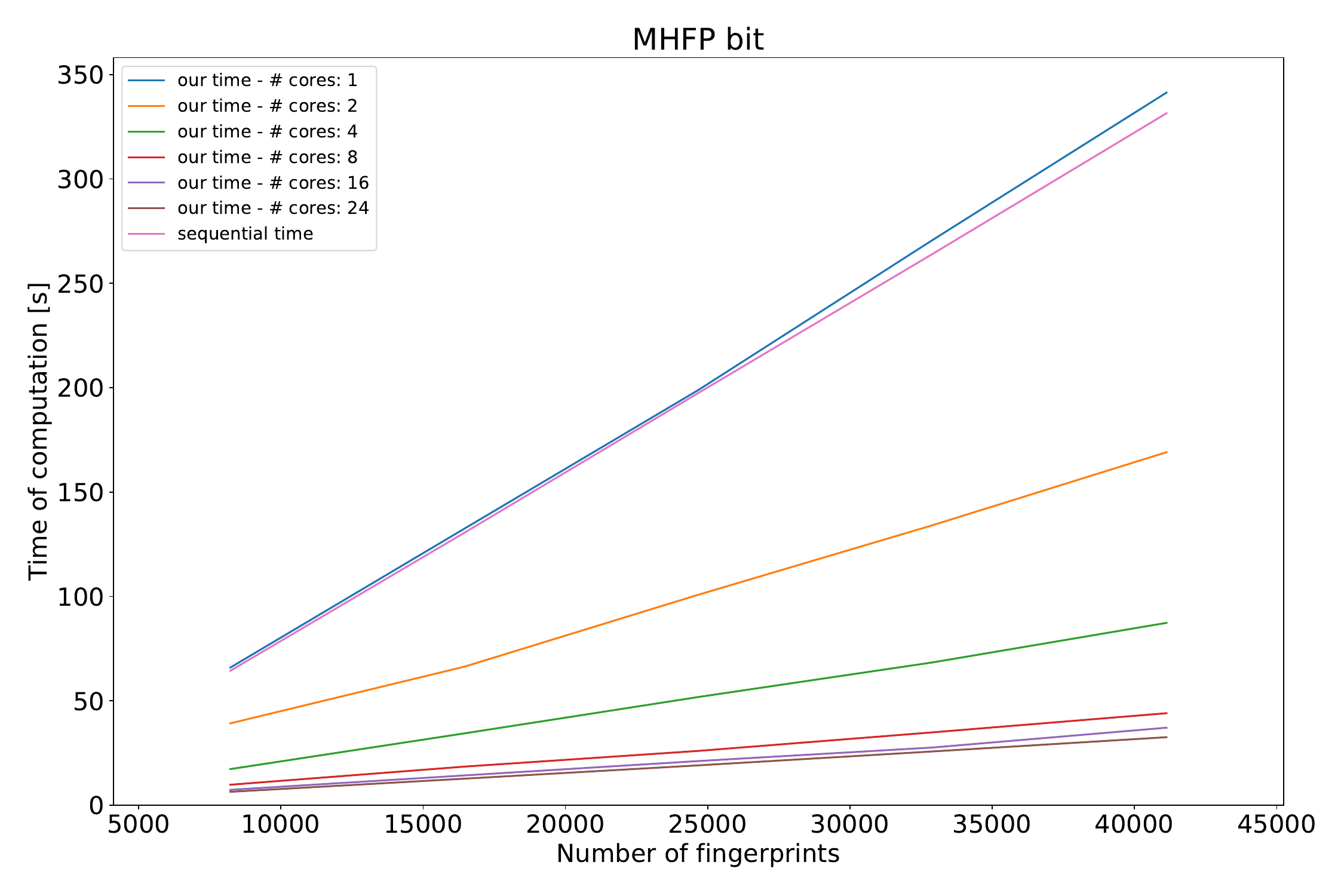}
  \centering
  \caption{Time results for MHFP Fingerprint — bit vector variant.}
  \label{fig:MHFP-bit}
\end{figure}

\begin{figure}[H]
  \centering
  \includegraphics[width=0.9\textwidth]{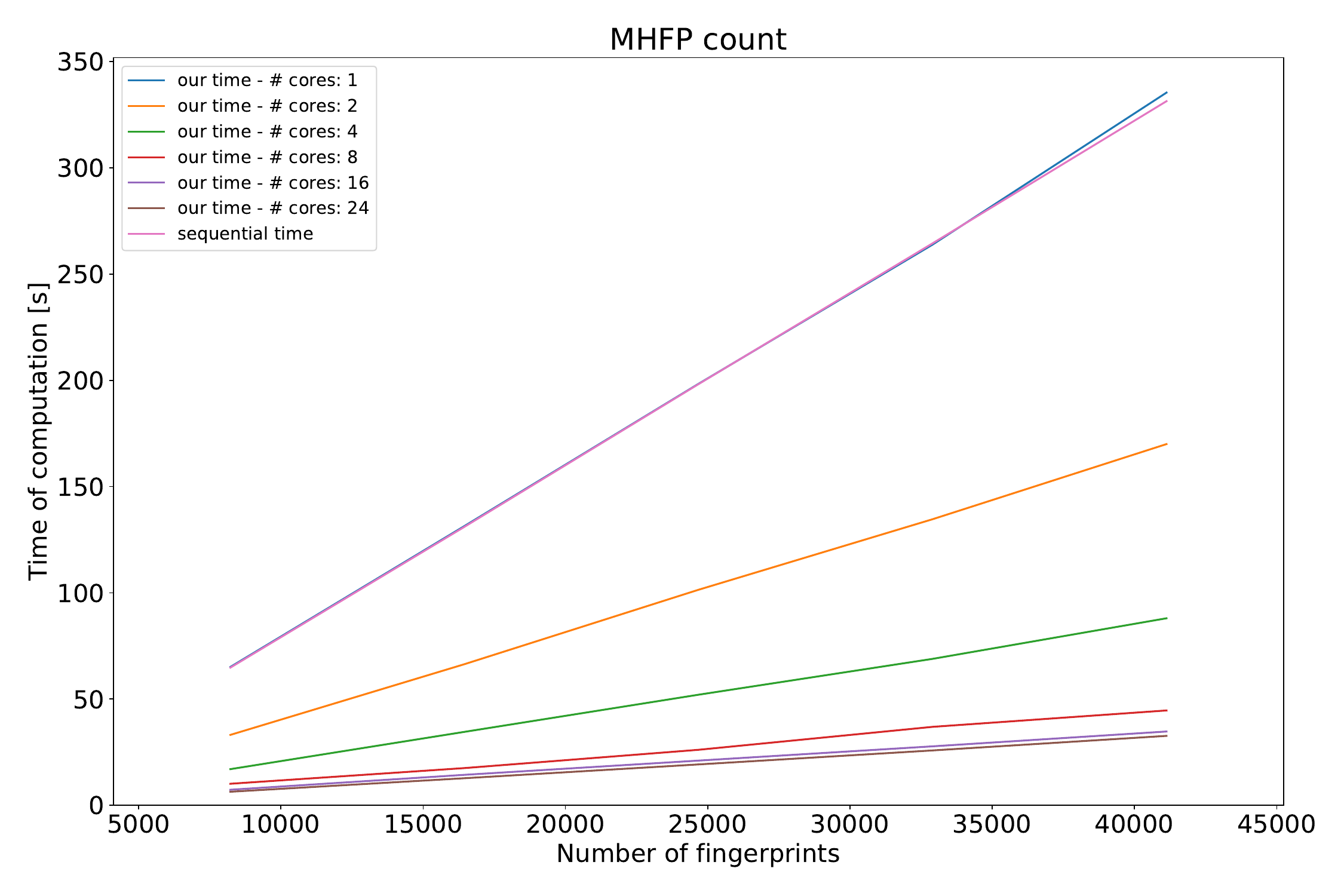}
  \centering
  \caption{Time results for MHFP Fingerprint — count vector variant.}
  \label{fig:MHFP-count}
\end{figure}

Additionally, we used the same dataset to perform molecular property prediction, described in \cref{sec:benchmark}. Using this dataset, we measured the computation time using an increasing number of processes, up to the number of cores available, each time increasing the number two times. Specifically, we checked 1, 2, 4, 8, 16 and 24 processes (24 was the total number of threads provided by the CPU of the machine used). In order to test scalability and speedup in relation to size of data, we performed the measurement on 20\%, 40\%, 60\%, 80\% and 100\% of molecules in the dataset. Each of those measurements has been performed 5 times, and we report average time for each data size. We observed a significant time performance improvement when using parallel processing for all fingerprints 
(\cref{fig:morgan-bit,fig:morgan-count,fig:atom-pair-bit,fig:atom-pair-count,fig:topological-torsion-bit,fig:topological-torsion-count,fig:maccs,fig:erg,fig:map4-bit,fig:map4-count,fig:MHFP-bit,fig:MHFP-count}). It is an expected result of our project, and it proves that it was successful in improving the performance of fingerprints calculation.

For some fingerprints, namely ECFP, Atom Pair and Topological Torsion (\cref{fig:morgan-bit,fig:morgan-count,fig:atom-pair-bit,fig:atom-pair-count,fig:topological-torsion-bit,fig:topological-torsion-count}) our implementation performed better than RDKit even for one thread. The reason for this performance improvement is most probably the effective memory management resulting from joining the computation results.

RDKit displays significant difference in time performance between bit (\cref{fig:morgan-bit,fig:atom-pair-bit,fig:topological-torsion-bit}) and count (\cref{fig:morgan-count,fig:atom-pair-count,fig:topological-torsion-count}) fingerprint type without our library. It was quite unexpected, as the bit vector should take less time to compute than the count one. Most probably RDKit algorithms for computing those two vector types differ highly in implementation and, therefore, in performance and this was the reason for the observed difference. This difference was not visible for sequential implementation of MAP4 and MHFP fingerprints (\cref{fig:map4-bit,fig:map4-count,fig:MHFP-bit,fig:MHFP-count}), as they were implemented without help of RDKit.

Out of RDKit fingerprints, the MACCS Keys (\cref{fig:maccs}) took the longest time to compute. As this is a fingerprint that searches for predefined features, the reason for that is most probably rooted in their complexity.

MAP4 and MHFP fingerprints took much longer to compute than other fingerprints. It was likely due to their implementation, which relies almost solely on Python. With MAP4 fingerprint taking over 200 seconds (\cref{fig:map4-bit,fig:map4-count}) and MHFP over 300 seconds  (\cref{fig:MHFP-bit,fig:MHFP-count}) on average to compute the whole HIV dataset with one thread, our parallel implementation is crucial to apply those fingerprints to larger datasets. The sequential processing is far too slow for performing machine learning, such as hyperparameter tuning, in reasonable time.

E3FP fingerprint proved to be too complex to compute in reasonable time even with our implementation. Computation time for the first one hundred molecules of the HIV dataset took over nine minutes. The reason for this was Python implementation and complexity of generating three-dimensional conformers. The resulting time is far too long to perform any reliable performance measurements. We plan to reimplement this fingerprint in the future using either Cython or C++ programming language to optimize the performance of sequential processing.

\clearpage


\clearpage

\clearpage

\clearpage

\clearpage

\clearpage

\clearpage

%%%%%%%%%%%%%%%%%%%%%%%%%%%%%%%%%%%%%%%%%%%%%%%%%%%%%%%%%%%%%%%%%%%%%%%%%%%%%%%

\section{Application to molecular property prediction}
\label{sec:benchmark}
\noindent We assessed the performance of each fingerprint for molecular property prediction. This is a machine learning (ML) classification task that uses fingerprint vectors as input features. To do this, we computed our fingerprints for three benchmark datasets from the MoleculeNet benchmark \cite{moleculenet}. It is commonly used for benchmarking prediction algorithms. Some important properties of the datasets are presented in \cref{table:three-datasets-properties}.

\begin{figure}[H]
  \centering
  \includegraphics[width=0.9\textwidth]{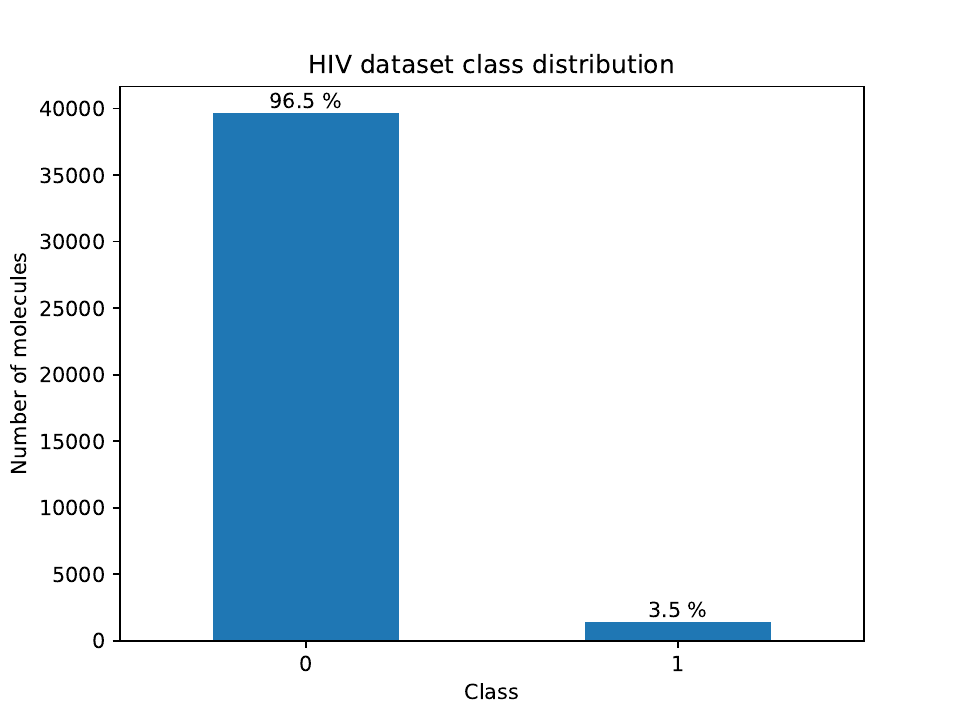}
  \centering
  \caption{Class distribution visualization for HIV dataset.}
  \label{fig:HIV-classes}
\end{figure}

For the first task, we use HIV benchmark dataset \cite{moleculenet-hiv}. The task is to predict whether a given molecule has a potential to be an HIV suppressor. It consists of molecules in SMILES string format and their binary class labels. The dataset displays significant class imbalance \cref{fig:HIV-classes}, since the positive class constitutes barely 3.5\% of all molecules in the dataset.

\begin{figure}[H]
  \centering
  \includegraphics[width=0.9\textwidth]{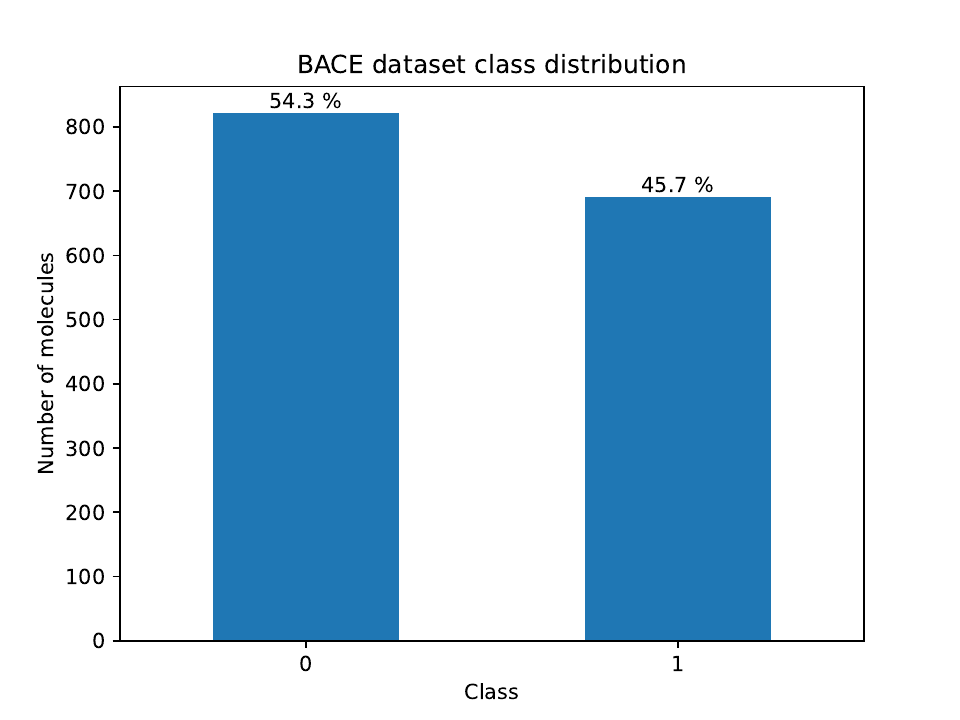}
  \centering
  \caption{Class distribution visualization for BACE dataset.}
  \label{fig:BACE-classes}
\end{figure}

The BACE dataset \cite{moleculenet-bace} provides binding results for a set of inhibitors of human beta-secretase 1 (BACE-1) — a protein enzyme playing a significant role in development of Alzheimer's disease. Classes in this dataset \cref{fig:BACE-classes} are much more balanced than in HIV.

\begin{figure}[H]
  \centering
  \includegraphics[width=0.9\textwidth]{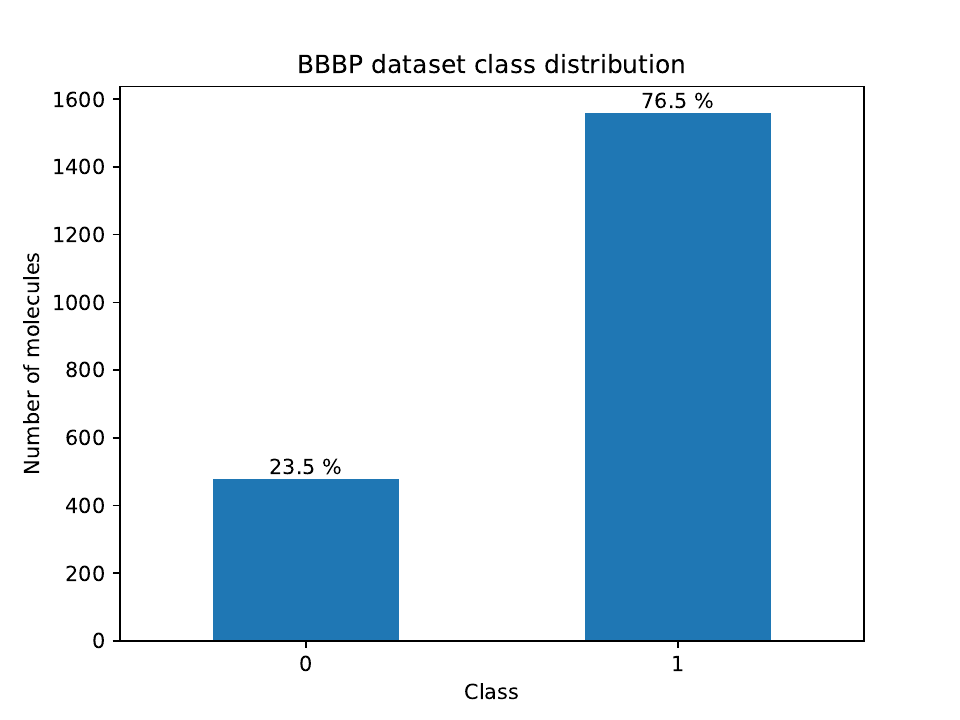}
  \centering
  \caption{Class distribution visualization for BBBP dataset.}
  \label{fig:BBBP-classes}
\end{figure}

BBBP — The BBBP (Blood-Brain Barrier Penetration dataset) \cite{moleculenet-bbbp} comes from a recent study on the modeling and prediction of the blood-brain barrier permeability \cite{moleculenet}. We observe opposite class balancing in this dataset, with the positive class being the majority, as illustrated in \cref{fig:BBBP-classes}. 

In the problem of molecular property prediction, the graph sizes are typically small. However, the datasets contain large numbers of them. This is a great example of use case, that benefits from parallel processing.

\begin{table}[H]
\begin{center}
\begin{tabular}{|c c c c c|} 
\hline
Dataset & \# Graphs & Avg. \# Nodes & Avg. \# Edges & \# Classes \\ [0.5ex] 
\hline\hline
HIV & 41127 & 25.5 & 27.5 & 2 \\
\hline
BACE & 1513 & 34.1 & 36.9 & 2 \\
\hline
BBBP &  2039 & 24.1 & 26.0 & 2 \\
\hline
\end{tabular}
\caption{The properties of datasets used for testing fingerprints in prediction tasks.}
\label{table:three-datasets-properties}
\end{center}
\end{table}

\clearpage

For each fingerprint, we trained three different classifiers and checked their Area Under Receiver Operating Characteristic curve (AUROC) score. This score has been chosen because of the imbalanced classification task. It was used in other works that use this dataset and the task \cite{moleculenet,ogb}. We used 80-10-10\% data split. However, due to usage of scaffold split (described below), the size of the test and validation sets was slightly higher.

The chemical space of molecules is very large. Because of that, machine learning solutions often perform significantly worse on previously unobserved parts of that space. If a model is trained on a randomly split dataset, there is a high possibility that the distribution of data will be very similar between training, testing, and validation set. In most ML solutions, this would be desirable. However, in chemoinformatics model evaluated with this kind of split would in reality perform worse during real world use. Such models usually perform prediction on data much different from the datasets they have been trained on. It is particularly important in de novo drug design, where new molecules are created in such a way that they are structurally different from the existing ones. This is done to increase efficiency or decrease side effects by using molecules with slightly different topology than the existing drugs. To mitigate this problem, we used a method called scaffold split \cite{moleculenet,bemis-murcko} for dividing the data into training, validation, and testing. Parts of the dataset selected for validation and testing represent a different distribution than one used for training. This way, we get closer to estimating the actual real-world performance of a model.

The trained classifiers were:
\begin{itemize} 
    \item logistic regression — linear probabilistic classifier that predicts a positive class probability based on a logistic transformation of a linear regression \cite{logistic-regression},
    \item Random Forest — a tree-based ensemble learning model that utilizes bagging based on bootstrap samples and random subspace methods \cite{random-forest},
    \item LightGBM — a tree-based ensemble learning model that utilizes gradient boosting \cite{lightgbm}.
\end{itemize}

The training was performed without hyperparameter tuning, using the simple default hyperparameters, in order to make sure that even simple models are able to yield satisfactory classification results. For fairness of comparison, we used bit (binary) variant for all fingerprints. Additionally, we train the models using balanced class weights (based on Scikit-learn settings), as the dataset displays a heavy imbalance \cref{fig:HIV-classes}.

The \cref{table:classification-rf,table:classification-logreg,table:classification-lgbm} show values of AUROC for each classifier trained with help of each of the fingerprints. For Random Forest Classifier, we perform the training using ten different random seeds and report mean and standard deviation of their scores. We do not do that for logistic regression and LightGBM as they do not consider random seed in their implementations.

\clearpage
\begin{table}[H]
\begin{center}
\begin{tabular}{|c c c|} 
\hline
Fingerprint & Test AUROC & Validation AUROC \\ [0.5ex] 
\hline
ECFP & 66.80 & 66.03\\  [1ex]
\hline
Atom Pair & 56.42 & 64.38\\  [1ex]
\hline
Topological Torsion & 66.20 & 70.38\\  [1ex]
\hline
MACCS Keys & 71.71 & 74.19\\  [1ex]
\hline
ErG & \textbf{72.82} & 75.27\\  [1ex]
\hline
MAP4 & 70.54 & 61.38\\  [1ex]
\hline
MHFP & 52.61 & 57.07\\  [1ex]
\hline
\end{tabular}
\caption{Classification scores on HIV dataset using logistic regression classifier.}
\label{table:classification-logreg}
\end{center}
\end{table}

\begin{table}[H]
\begin{center}
\begin{tabular}{|c c c|} 
\hline
Fingerprint & Test AUROC & Validation AUROC \\ [0.5ex] 
\hline
ECFP & 76.69 ± 1.26 & 83.06 ± 1.42\\  [1ex]
\hline
Atom Pair & \textbf{79.47 ± 1.00} & 80.80 ± 1.58\\  [1ex]
\hline
Topological Torsion & 76.75 ± 1.02 & 80.39 ± 1.59\\  [1ex]
\hline
MACCS Keys & 75.08 ± 1.55 & 81.10 ± 2.08\\  [1ex]
\hline
ErG & 74.38 ± 2.06 & 77.68 ± 2.50\\  [1ex]
\hline
MAP4 & 66.13 ± 1.30 & 64.33 ± 3.56\\  [1ex]
\hline
MHFP & 65.29 ± 1.41 & 66.40 ± 1.94\\  [1ex]
\hline
\end{tabular}
\caption{Classification scores on HIV dataset using Random Forest classifier.}
\label{table:classification-rf}
\end{center}
\end{table}

\begin{table}[H]
\begin{center}
\begin{tabular}{|c c c |} 
\hline
Fingerprint & Test AUROC & Validation AUROC \\ [0.5ex] 
\hline
ECFP & 74.62 & 70.74\\  [1ex]
\hline
Atom Pair & 73.15 & 77.39\\  [1ex]
\hline
Topological Torsion & 72.38 & 79.64\\  [1ex]
\hline
MACCS Keys & \textbf{74.95} & 76.97\\  [1ex]
\hline
ErG & 74.03 & 72.73\\  [1ex]
\hline
MAP4 & 66.58 & 65.13\\  [1ex]
\hline
MHFP & 54.23 & 58.22\\  [1ex]
\hline
\end{tabular}
\caption{Classification scores on HIV dataset using LightGBM classifier.}
\label{table:classification-lgbm}
\end{center}
\end{table}

\clearpage
The results show that all implemented fingerprints perform significantly better for Random Forest and LightGBM than they do for logistic regression. Better performance of tree-based learning algorithms over logistic regression indicates that the underlying data has more complex and non-linear relationships.

Our results compare very well with other classification algorithms published at the moment of writing this work.
Our best result — Random Forest with Atom Pair fingerprint — achieves 15th place on the official leaderboard with score 79.47 ± 1.00. This is especially significant, considering that we performed no hyperparameter tuning and this is just a baseline result for fingerprint-based approaches 
\cite{ogb,OGB-molhiv-ranking}. Other comparable solutions' results are presented in the Table \cref{table:ranking}.

\begin{table}[H]
\begin{center}
\begin{tabular}{|c c c|} 
 \hline
Ranking position & Classification method & Test AUROC score \\ [1ex]
\hline
13	& directional GSN \cite{directional-GSN} &	80.39 ± 0.90 \\ [1ex]
\hline 
14	& DGN \cite{DGN} & 79.70 ± 0.97  \\ [1ex]
\hline 
15	& DeeperGCN+FLAG \cite{DeeperGCN+FLAG} & 79.42 ± 1.20 \\ [1ex]
\hline 
16	& PHC-GNN \cite{PHC-GNN} & 79.34 ± 1.16 \\ [1ex]
\hline 
17	& PNA \cite{PNA} & 79.05 ± 1.32 \\ [1ex]
\hline

\end{tabular}
\caption{HIV benchmark leaderboard — competing solutions scores.}
\label{table:ranking}
\end{center}
\end{table}

Additionally, it came as a surprise that the MAP4 and MHFP fingerprints resulted in worse score than other fingerprints despite being newer, more modern and generally more advanced. This observation requires further testing on more molecular property prediction tasks, in order to establish whether they are generally useful for those types of applications.

\clearpage

We performed similar benchmarking for BBBP and BACE to see how well we can perform on data from different distribution. By performing more than one task, we verify the consistency of our solution and study the performance of our solution in different real-life scenario.

\begin{table}[H]
\begin{center}
\begin{tabular}{|c c c c|} 
\hline
Fingerprint & logistic regression & Random Forest & LightGBM [\%] \\ [0.5ex] 
\hline
ECFP & 66.80 & 76.69 ± 1.26 & 74.62\\ [1ex]
\hline
Atom Pair & 56.42 & \textbf{79.47 ± 1.00} & 73.15\\ [1ex]
\hline
Topological Torsion & 66.20 & 76.75 ± 1.02 & 72.38\\ [1ex]
\hline
MACCS Keys & 71.71 & 75.08 ± 1.55 & 74.95\\ [1ex]
\hline
ErG & 72.82 & 74.38 ± 2.06 & 74.03\\ [1ex]
\hline
MAP4 & 70.54 & 66.13 ± 1.30 & 66.58\\ [1ex]
\hline
MHFP & 52.61 & 65.29 ± 1.41 & 54.23\\ [1ex]
\hline
\end{tabular}
\caption{Test AUROC scores for HIV dataset.}
\label{table:test-results-HIV}
\end{center}
\end{table}

\begin{table}[H]
\begin{center}
\begin{tabular}{|c c c c|} 
\hline
Fingerprint & logistic regression & Random Forest & LightGBM \\ [0.5ex] 
\hline
ECFP & 69.04 & 82.00 ± 0.82 & 80.32\\ [1ex]
\hline
Atom Pair & 80.31 & 85.19 ± 1.07 & 84.59\\ [1ex]
\hline
Topological Torsion & 69.23 & \textbf{85.58 ± 0.79}  & 81.13\\ [1ex]
\hline
MACCS Keys & 74.44 & 79.03 ± 1.15 & 76.00\\ [1ex]
\hline
ErG & 66.24 & 79.45 ± 0.95 & 78.60\\ [1ex]
\hline
MAP4 & 44.01 & 58.91 ± 3.50 & 63.92\\ [1ex]
\hline
MHFP & 50.10 & 58.58 ± 4.09 & 55.12\\ [1ex]
\hline
\end{tabular}
\caption{Test AUROC scores for BACE dataset.}
\label{table:test-results-bace}
\end{center}
\end{table}

\begin{table}[H]
\begin{center}
\begin{tabular}{|c c c c|} 
\hline
Fingerprint & logistic regression & Random Forest & LightGBM \\ [0.5ex] 
\hline
ECFP & 58.00 & 67.48 ± 0.97 & 60.26\\ [1ex]
\hline
Atom Pair & 61.02 & \textbf{71.76 ± 0.82} & 68.55\\ [1ex]
\hline
Topological Torsion & 57.20 & 66.23 ± 0.91 & 63.41\\ [1ex]
\hline
MACCS Keys & 71.02 & 69.56 ± 0.78 & 68.17\\ [1ex]
\hline
ErG & 53.41 & 71.42 ± 0.82 & 71.52\\ [1ex]
\hline
MAP4 & 61.00 & 59.54 ± 1.54 & 59.23\\ [1ex]
\hline
MHFP & 53.16 & 56.19 ± 3.50 & 57.38\\ [1ex]
\hline
\end{tabular}
\caption{Test AUROC scores for BBBP dataset.}
\label{table:test-results-bbbp}
\end{center}
\end{table}

We observe that Atom Pair fingerprint performs very well for all three classification problems \cref{table:test-results-HIV,table:test-results-bbbp,table:test-results-bace}. This is an important observation, since this fingerprint was not commonly used by researchers in this field. For BACE dataset, the best result was achieved with closely related Topological Torsion fingerprint, with only a small advantage over Atom Pair. Random Forest  proved to be the best model for this task out of the three we used.

The results compare very well to solutions based on various Graph Neural Networks (GNNs), including pretrained models \cite{pretraining-GNNs}.
Their scores are presented in \cref{table:scores-for-pretraining-gnns}. The best GNN result and the best result of our approaches were marked in bold. We selected two best fingerprints and Random Forest classifier for this comparison. We outperform all GNNs variants on BACE and BBBP, and come very close on HIV. This shows that molecular fingerprints are still very important nowadays in molecular property prediction.

\begin{table}[H]
\begin{center}
\begin{tabular}{|c c c c c|} 
\hline
Model architecture & Pretrained? & HIV & BACE &  BBBP \\ [0.5ex] 
\hline
GIN & No &  75.30 ± 1.90 &  70.1 ±5.4 & 65.8 ± 4.5\\ [1ex]
\hline
GIN & Yes & \textbf{79.9 ± 0.7} &  \textbf{84.5 ± 0.7}  & 68.7 ± 1.3 \\ [1ex]
\hline
GCN & No & 75.7 ± 1.1 & 73.6 ± 3.0  & 64.9 ± 3.0 \\ [1ex]
\hline
GCN & Yes & 78.2 ± 0.6  & 82.3 ± 3.4  & \textbf{70.6 ± 1.6} \\ [1ex]
\hline
GraphSAGE & No & 74.4 ± 0.7  & 72.5 ± 1.9  &  69.6 ± 1.9 \\ [1ex]
\hline
GraphSAGE & Yes & 76.2 ± 1.1  & 80.7 ± 0.9  & 63.9 ± 2.1 \\ [1ex]
\hline
GAT & No &  72.9 ± 1.8 & 69.7 ± 6.4  & 66.2 ± 2.6 \\ [1ex]
\hline
GAT & Yes & 62.5 ± 1.6  &  64.3 ± 1.1 &  59.4 ± 0.5 \\ [1ex]
\hline
\hline
Atom Pair + RF & N/A &  \textbf{79.5 ± 1.0} & 85.2 ± 1.0  & \textbf{71.8 ± 0.8} \\ [1ex]
\hline
Topological Torsion + RF & N/A & 76.8 ± 1.0  &  \textbf{85.6 ± 0.8} &  66.2 ± 0.9 \\ [1ex]
\hline
\end{tabular}
\caption{The scores of various GNNs, compared to our best fingerprint and classifier.}
\label{table:scores-for-pretraining-gnns}
\end{center}
\end{table}

Those experiments were intended to be a simple assertion of the product quality. It would be interesting to further explore the library's possibilities by performing similar training on different datasets and with hyperparameter tuning. However, as conducting research using our library is not the main part of this project, we decided to focus on implementation aspects. However, we note that possibility as a future work.

%%%%%%%%%%%%%%%%%%%%%%%%%%%%%%%%%%%%%%%%%%%%%%%%%%%%%%%%%%%%%%%%%%%%%%%%%%%%%%%

\chapter{\ChapterTitleWorkOrganization}
\label{sec:organizacja-prac}

\noindent In this chapter, we describe the software development process and project organizational details.

\section{Division of work}
\label{sec:division-of-work}

\noindent
Michał:
\begin{itemize}
    \item Created initial draft of library interface and architecture
    \item Implemented majority of fingerprints
    \item Incorporated parallelism using \texttt{joblib}
    \item Created proof-of-concept notebook for molecular property prediction
    \item Added logging, including integration with RDKit logging mechanism
    \item Wrote \cref{sec:cel-wizja,sec:functional-scope} and parts of \cref{sec:division-of-work} of this work.
\end{itemize}

\noindent
Przemysław:
\begin{itemize}
    \item Poetry setup with dependency management
    \item Made library pip-installable and released the library on PyPI
    \item Developed CI/CD and GitHub Actions
    \item Wrote \cref{sec:organizacja-prac,sec:devops} and parts of \cref{sec:quality-and-maintenance} of this work.
\end{itemize}

\noindent
Piotr:
\begin{itemize}
    \item Created benchmarking of efficiency and time of algorithms. 
    \item Implemented example use case of molecular property prediction, including comparison with other solutions.
    \item Managed work division and organization
    \item Wrote unit tests with extended variety of tested molecules and debug of algorithms
    \item Implemented various Python-related optimizations
    \item Implemented and testing of scikit-learn compatibility elements
    \item Refactored fingerprints to use RDKit generator interface
    \item Added sparse matrices functionality
    \item Created pre-commit hooks
    \item Created MoSCoW prioritization for the project \cref{sec:moscow}
    \item Wrote \cref{sec:wyniki,sec:conclusion,sec:biblioteka,sec:fingerprints,sec:efficiency,sec:testing} and parts of \cref{sec:division-of-work} of this work.
\end{itemize}

\section{Project management and utilized tools}
\noindent To manage our project, we used Agile oriented techniques like Scrum. We designed a Scrum board on Trello platform in order to organize our work and separate its stages until the product was finished. We used Git and GitHub for version control. Every week we tried to iteratively deliver new functionalities following our previously created MoSCoW prioritization of features. We constantly stayed in touch through online chats and meetings on platforms like Microsoft Teams, Google Meet, Discord, and Messenger in order to be able to share the knowledge on new features and get help we needed to deliver the project on time. Once the project basis was done, we switched to GitHub Issues to integrate all the workflow into one place and simplify the process of developing the product. Further features and bugs were brought up as new issues that we could all see and decide which of them we wanted to work on in our weekly iterations. Every feature development involved creating and reviewing new branches and pull requests, later were integrated into main branch.

\section{Stages of work}
\begin{enumerate}
    \item \textit{March}: defined the scope of our work, focused mostly on organization. Created a Scrum board and a proof-of-concept notebook. Familiarized ourselves with RDKit environment and existing solutions for fingerprint calculation.
    \item \textit{April}: implemented interface for fingerprints, basic unit tests, ECFP, MACCS Keys, Atom Pair, Topological Torsion and ERG fingerprints. Basic unit tests were added.
    \item \textit{May}: added multiprocessing with Joblib. Several reviews and corrections of code. Refactored code, introduced abstractions to reduce code duplication. Finished writing the outline of our work and mostly completed the first chapter.
    \item \textit{June}: added optional result types. Refactored code in fingerprint classes to work properly with scikit-learn.
    \item \textit{July}: Code refactoring and bugfixes after further extended testing.
    \item \textit{August}: implemented RDKit generator interface for calculation of selected fingerprints. Integrated Poetry into the project.
    \item \textit{September}: added a time benchmarking for ECFP, MACCS Keys, Atom Pair, Topological Torsion and ERG fingerprints. Drafted \texttt{pyproject.toml} file to manage dependencies.
    \item \textit{October}: implemented MAP4, MHFP and E3FP fingerprints from scratch. Configured \texttt{pytest} and PyPI platforms to integrate our library.
    \item \textit{November}: MAP4, MHFP and E3FP were added to benchmark, several reviews and corrections to these fingerprints were made, and minor fixes were done. Refactored the repository for compatibility with PyPI, updated dependencies. Settled legal aspects of project publication, added rules and instructions for future users and developers.
    \item \textit{December}: Final version of benchmark. Cleanup of dependencies. Implemented example use case of molecular property prediction. Published the project on PyPI. Integrated CI/CD and code standardization features. Finished writing our work — completed fingerprints descriptions, added tables and figures, and described the process of our work.
\end{enumerate}

%%%%%%%%%%%%%%%%%%%%%%%%%%%%%%%%%%%%%%%%%%%%%%%%%%%%%%%%%%%%%%%%%%%%%%%%%%%%%%%
\chapter{\ConclusionAndFutureWork}
\label{sec:conclusion}

\noindent This chapter describes our plans for future work with the project and explores possibilities of putting the \texttt{scikit-fingerprints} library to use. It describes a possible work impact of the created product such as conferences and publications and includes a summary of the project.

\section{Future work}
\label{subsec:future-work}
The library is already a feature-rich product, however, the field of chemoinformatics is vast and there are more complex molecular fingerprints. We can implement more of the traditional fingerprints, such as PubChem descriptors, and deep learning based nodes, like CDDD \cite{cddd} or Mol2Vec \cite{mol2vec}. We took notes of such plans for future work as issues in our GitHub repository.

Our project is focused on computing molecular fingerprints, but there are many other useful tools that could benefit from parallelism. The approach to parallel processing used in our library can be utilized to implement an easier way for transforming molecules from or to SMILES string format or standardized them.

\section{Work impact}
Our project was presented at ELEMENTS 2023 conference at AGH University of Science and Technology. We additionally want to present it during other ML or chemoinformatics conferences, in order to maximize public knowledge of this library.
We want to publish the effects of our effort in science journals in the future work and submit a paper in the first quarter of the 2024.

The delivered library can be used for scientific research that want to work on in the future. It opens a door for research in the field of chemoinformatics that will be more complex than before. Our plans are to use this tool to conduct research oriented around molecular property prediction problem. The library allows us to efficiently do that using fingerprints that, for this particular task, have not been used before, e.g. by using so-called compound or complex fingerprints, using concatenations of multiple fingerprints and tuning the whole ensemble at once. This might allow us to continue the development and use of our library during our Master's degree or PhD studies.

\clearpage

\section{Conclusion and summary}
We consider our project to be successful. We managed to deliver a highly functional library that is compatible with \texttt{scikit-learn}, is fast, easy to install and intuitive. It is a great tool, allows users to incorporate molecular fingerprints into their machine learning workflow with ease. The repository uses high code quality and CI/CD tools that help maintain good repository structure and allow developers to contribute to the project and maintain it in the future.

\clearpage

\phantomsection
\addcontentsline{toc}{chapter}{Bibliography}
\printbibliography
\end{document}